\documentclass[]{elsarticle}

\usepackage{amssymb}  
\usepackage{amsthm}   
\usepackage{latexsym} 
\usepackage{mathtools} 
\usepackage{siunitx}
\usepackage{lineno}
\usepackage{caption}
\usepackage{subfig}
\usepackage{color}

\begin{document}

\begin{frontmatter}

\title{Vertex Fitting In Low-Material Budget Pixel Detectors}

\author{Andrea Loreti}

\affiliation{organization={Department of Physics, University of Liverpool},
            addressline={The Oliver Lodge Laboratory}, 
            city={Liverpool},
            postcode={L69 7ZE}, 
            state={},
            country={United Kingdom}}

\begin{abstract}
This paper provides a detailed description of a vertex fitting algorithm designed for precision measurements in low-energy particle physics experiments. An accurate reconstruction of low-momentum trajectories is facilitated by reducing the material budget of the detector to a few per mill of the radiation length. This decreases the multiple scattering experienced by particles within the detector, thereby enhancing vertex fitting accuracy. However, in the case of light detection systems, the intrinsic spatial resolution of the sensors imposes further constraints on the final vertex resolution that require careful consideration. The algorithm developed in this study addresses both multiple scattering and spatial resolution aspects in the context of vertex fitting, specifically tailored for light pixel detectors. Furthermore, this works presents a detailed examination of the vertex reconstruction within the low-material budget pixel detector of the Mu3e experiment.

\end{abstract}


\end{frontmatter}

\section{Introduction}\label{Intro}
With the increase of the instantaneous and integrated beam luminosities, the requirements of particle physics experiments for precise tracking and vertexing detectors, with high radiation tolerance, have become more stringent e.g., \cite{SNOEYS2023168678,Hartmut2018,CARNESECCHI2019608,MOSER201685}. In this regard, silicon pixel sensors can provide high granularity, low material budget structures and the radiation-hardness that most experiments need, e.g., \cite{MOSER201685,SPANNAGEL2019612,Abelev_2014,ARNDT2021165679}. It is important, however, that precise detection systems are developed in conjunction with equally performing analysis methods for the reconstruction of particle trajectories and decay vertices, e.g. \cite{FRUHWIRTH1987444,BILLOIR1992139,Waltenberger2007,RevModPhys.82.1419}. To this aim, several fitting algorithms have been designed and optimized over the years to deal with hit selection, pattern recognition, errors calculations and high track multiplicity (see for instance \cite{RevModPhys.82.1419,Mankel_2004} and references therein). The practical implementation of these methods must be tailored around the actual detector and magnetic field configuration of the experiments. This makes tracking and vertexing a topic which is in continuous evolution adapting itself to the new challenges set by upcoming experiments.

This study addresses the problem of vertex fitting in the low-material budget pixel detector of the Mu3e experiment \cite{ARNDT2021165679}. As explained in section \ref{Tracker}, the detector design has been optimized to minimize the effects of Multiple Coulomb Scattering (MS) on particle trajectories and signal kinematics. 
However, in the case of light detectors like Mu3e, the intrinsic pixel resolution emerges as an additional constraint in the vertex reconstruction demanding careful consideration. The present vertex fitting takes into account both MS and pixel resolution errors, alongside any uncertainties derived from energy losses within the detector. Section \ref{Vertexing} provides a detailed description of the algorithm whilst in section \ref{Mu3eCaseStudy}, a comparative study is conducted among different inclusive scenarios that encompass: (A) MS only; (B) MS and pixel resolution together; (C) all sources of errors (MS, pixel resolution and energy losses) are included. 

\section{The Mu3e low-material budget pixel detector}\label{Tracker}
\begin{figure*}[h!]
\centering
\includegraphics[scale=0.4]{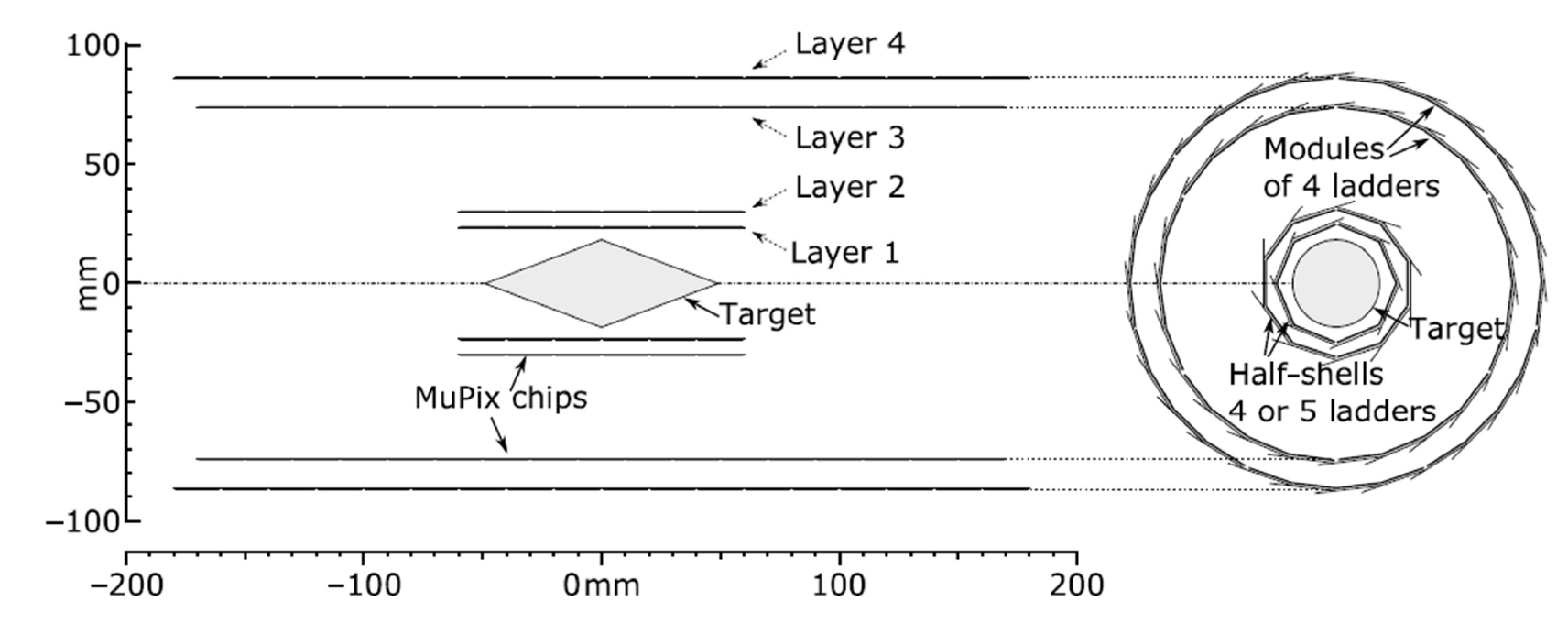}\caption{Scheme of the central station of the Mu3e pixel detector, side and x-y views \cite{ARNDT2021165679}.}\label{Mu3eDetector}
\end{figure*}
The Mu3e experiment aims to find or exclude the rare Charged Lepton Flavour (CLF) violating muon decay:
\begin{equation}\label{Mu3eDecay}
\mu^+\rightarrow e^+e^-e^+
\end{equation} 
at final Branching Ratios (BR) $>10^{-16}$ \cite{ARNDT2021165679}. The first phase of the experiment is designed to use the $\pi$E5 low-energy muon source ($10^8\,$Hz) at the Paul Scherrer Institute, where Mu3e is under commissioning. The aimed single event sensitivity of the experiment is $4$ orders of magnitude higher than previous experimental upper limits (BR $=1.0\cdot 10^{-12}$) \cite{BELLGARDT19881} and $38$ orders of magnitude lower than theoretical Standard Model (SM) calculations (BR $=10^{-54}$), e.g., \cite{MARCIANO1977303,RevModPhys.73.151}. However, new theoretical studies predict the existence of additional degrees of freedom beyond the SM. These extensions may lead to CLF violation potentially detectable by upcoming experiments such as Mu3e, e.g., \cite{KAKIZAKI2003210,DEGOUVEA201375}. Consequently, the observation of $\mu^+\rightarrow e^+e^-e^+$ at sensitivity levels achievable in the Mu3e experiment will be a clear signature of new physics beyond the SM \cite{DEGOUVEA201375,BERNSTEIN201327}. 

The process in (\ref{Mu3eDecay}) yields a relatively simple decay topology with the 3 final state leptons produced at the same vertex of interaction and momentum vectors, $\vec{p}$, determined by the energy and momentum conservation for decaying muons at rest. The main background processes in Mu3e measurements are the muon internal conversion $\mu^+\rightarrow e^+e^-e^+ + \nu_{e} + \bar{\nu}_{\mu}$ (BR $\approx 10^{-5}$) and the combination of one electron and 2 positrons from independent sources, e.g., one Bhabha electron plus two Michel positrons $\mu^+\rightarrow 2\times(e^+ + \nu_e + \bar{\nu}_{\mu}) + e^-$ \cite{BELLGARDT19881}. The aimed single event sensitivity of $2\cdot10^{-15}$, during phase I of the experiment, can be achieved with an energy-momentum resolution of $\lesssim 1$\,MeV and by using precise vertexing and timing systems \cite{ARNDT2021165679}.

The energy spectrum of the decay particles in the Mu3e experiment extends up to $m_{\mu}/2$, where $m_{\mu}$ is the muon mass. In this low-energy region, MS poses a serious challenge to the reconstruction of particle trajectories and signal kinematics. To minimize MS, Mu3e uses a low-material budget pixel detector ($0.1\%$ of the radiation length, X$_o$, per layer). This is made of high-voltage monolithic active pixel sensors \cite{PERIC2007876} thinned down to \SI{50}{\micro\meter} or $0.05\%$X$_o$. The rest of the material budget of the detector is used in the flex-tape that provides mechanical support and the electrical routing to the sensors. 

Figure \ref{Mu3eDetector} shows the schematic of the foreseen Mu3e tracker central station which is important for vertex fitting and track reconstruction \cite{ARNDT2021165679}. Two re-curl stations, one up-stream and one down-stream, will also be part of the final detector design. These increase the angular acceptance of the experiment and allow to measure long-armed trajectories to achieve improved momentum resolution. 

The layers have cylindrical symmetry and are concentrically placed around the target, a hollow double cone made of Mylar 100\,mm in length and with a base radius of 19\,mm. The target is placed in a solenoid magnetic field of 1\,T with the base at a minimum distance of $\approx4$\,mm from the innermost layer of the pixel tracker. Particle trajectories bend inwards following helical trajectories around the field lines possibly making multiple re-curls. Each layer of the pixel detector is sectioned in sub-elements called ladders. A ladder is a series of chips mounted on the same flex-tape. There are 8, 10, 24 and 28 ladders for layer 1, 2, 3 and 4, respectively. For instance, the innermost layer of the tracker, crucial for vertex fitting, is made of 8 ladders each one tilted by a 45$^{\circ}$ angle with respect to the neighbours.  This configuration forms a 8-sided surface which extends for $\sim12$\,cm or 6 chips length, see figure \ref{Mu3eDetector}.

The intrinsic detector spatial resolution is set by the pixel sensitive area, $80\times80$\SI{}{\micro\meter}$^2$.

\section{Vertex fitting in the Mu3e pixel detector}\label{Vertexing}
Vertex reconstruction can be accomplished in two steps: vertex finding and vertex fitting, see e.g., \cite{RevModPhys.82.1419}. The former consists in grouping trajectories that have been most likely produced in the same decay process. The latter involves finding the most likely vertex coordinates $(x,y,z)_\text{v}$ and the initial momentum vectors of all clustered tracks. In Mu3e, vertex finding is accomplished by considering all possible combinations of two positive and one negative tracks in the detector within time frames of 64\,ns. For the vertex fitting, a least-squares optimization algorithm has been developed based on the method illustrated in \cite{BILLOIR1992139}.

\subsection{Track parameters and uncertainties}\label{params_uncert}
Particle trajectories are defined by 6 parameters ($[x,y],z,\phi,\lambda,k)$. These are the coordinates of one point along the track, the angles $\phi$ and $\lambda$ defining the direction of the tangent vector to the trajectory and the factor $k=(p/q)^{-1}$ where $q$ is the charge of the particle and $p$ is the magnitude of the momentum vector. The following relationships among the momentum components in the global Cartesian frame and the angles $\phi$ and $\lambda$ hold true:
\begin{equation}\label{initialSystem}
\left\lbrace
\begin{aligned}
p_x &= p\, \text{cos}(\lambda)\, \text{cos}(\phi)\, , \\
p_y &= p\, \text{cos}(\lambda)\, \text{sin}(\phi)\, ,\\
p_z &= p\, \text{sin}(\lambda)\, ,\\
R_{\perp} & = \frac{p\,\text{cos}(\lambda) }{q\,B} \, ,\\
\phi&\in[0,2\pi]\, ,\\
\lambda&\in[-\pi/2,\pi/2]\,,
\end{aligned}
\right.
\end{equation}
where $B=B_z$ is the homogeneous magnetic field directed along the beam-line $\hat{z}$ and $R_{\perp}$ is the transverse radius of the trajectory. 

\begin{figure}[th!]
\centering
\includegraphics[scale=0.7]{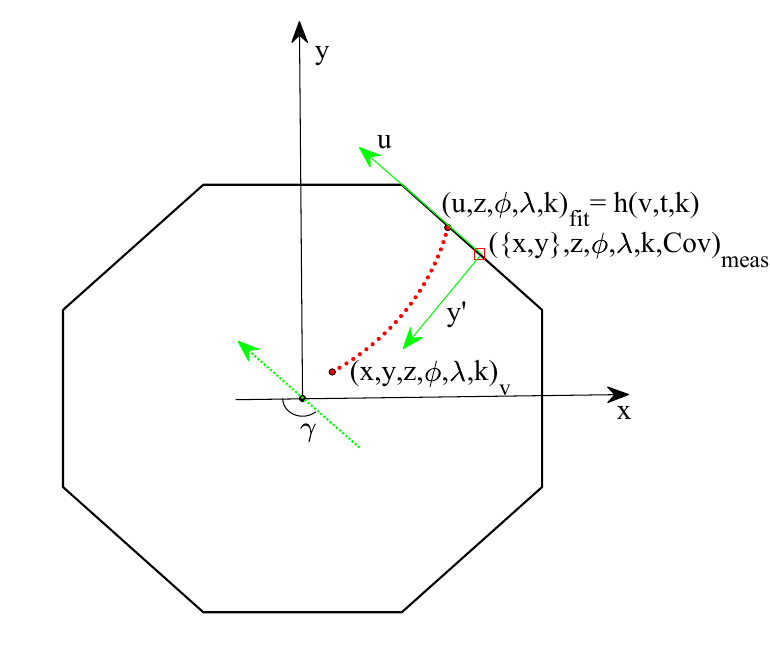}
\caption{Sketch of the RS considered in this study. The axes of the global (local) reference frame are drawn in black (green); the $\hat{z'}$ axis perpendicular to the $x-y$ plane is not shown. The parameters (${x,y},z)_{\text{meas}}$ are placed in the geometrical centre of the pixel surface \textcolor{red}{$\square$}. The fit parameters ($u,z,\phi,\lambda,k)_{\text{fit}}$ are obtained by forward propagating vertex parameters ($x,y,z,\phi,\lambda,k)_{\text{v}}$ via the function $h=h(\textbf{v}=(x,y,z)_\text{v},\textbf{t}(\phi_\text{v},\lambda_\text{v}),k_\text{v}))$. }\label{LocalRefFrame}
\end{figure}

In the present study, track parameters at a given Reference Surface (RS) along with their covariance matrix ($\Xi$): ($[x,y],z,\phi,\lambda,k,\Xi)_{\text{meas}}$ are considered as input measurements for the fit. The RS is the innermost layer of the pixel detector which is the closest one to the expected real vertex position. The coordinates $x,y$ are not independent and can be given in a single expression by using a local reference frame with center in the middle of the pixel area $(\bar{x}, \bar{y}, \bar{z})$ and base vectors $(\hat{u},\hat{z'})$. The vector $\hat{z'}$ is parallel to the global coordinate $\hat{z}$ while $\hat{u}$ is perpendicular to it and parallel to the pixel surface, see figure \ref{LocalRefFrame}. The equations that link the local coordinates to the global ones are:
\begin{equation}\label{EqTransform}
\begin{array}{lll}
u=\left( x - \bar{x}\right) \text{cos}(\gamma) + \left( y - \bar{y}\right)\text{sin}(\gamma)\, ,\\
z' = z -\bar{z}\, ,\\
\phi'=\phi\, ,\\
\lambda'=\lambda\,  \\
k'=k\,
\end{array} 
\end{equation}
where the angle $\gamma$ is the orientation of the pixel with respect to the $x$ axis of the global reference frame. Following the equations in \ref{EqTransform}, the track parameters at the RS become $(u,z,\phi,\lambda,k)$ \footnote{The coordinate $z'$ can be replaced by $z$ given that the constant $\bar{z}$ in eq. \ref{EqTransform} does not contribute to the calculations of the derivatives and residuals carried out in section \ref{least-square}}.

The covariance matrix can be written as the sum of two terms, i.e., $\Xi_{\text{meas}} = \Xi_{\text{track}} + \Xi_{\text{RS}}$. The term  $\Xi_{\text{track}}$ accounts for the uncertainties accumulated during track fitting while $\Xi_{\text{RS}}$ accounts for MS and  pixel resolution at the RS. Pixel resolution contributes to the smearing of the hit position by $\sigma_l = l/\sqrt{12}$, where $l$ is the length of the pixel side. The MS changes the direction of the track upon crossing the RS. The tilt can be approximated by a Gaussian distribution with zero mean and standard deviation $\theta_{\text{MS}}$ \cite{Yao_2006}:
\begin{equation}\label{MS}
\theta_{\text{MS}} = 13.6\,\left[ \text{MeV/c}\right] \,\frac{q}{p}\sqrt{\frac{d}{\beta^2\text{X}_o}}\left(1+0.038\,\text{ln}\left(\frac{d}{\text{X}_o}\right)  \right)\, .
\end{equation}
In the previous equation $\beta$ is the relativistic velocity and $d$ is the distance travelled by the particle in the RS. From eq. \ref{MS}, the changes in $\phi$ and $\lambda$ due to MS, $\sigma_{\phi}$ and $\sigma_{\lambda}$, can be obtained by projecting $\theta_{\text{MS}}$ onto the transverse and longitudinal planes, respectively, see e.g., \cite{VALENTAN2009728}:
\begin{equation}\label{EqsigmaLam}
\sigma_{\lambda} = \theta_{\text{MS}}\, ,\\
\end{equation}
and
\begin{equation}\label{EqsigmaPhi}
\sigma_{\phi} = \theta_{\text{MS}}/\text{cos}(\lambda)\,  .
\end{equation}
In addition to MS and pixel resolution, an error $\sigma_k$ on the track curvature (or $k$) is introduced in $\Xi_{\text{meas}}$ to account for the energy lost by electrons and positrons in the detector, see e.g., \cite{Mankel_2004}. Although energy losses in the Mu3e RS are negligible, they are nevertheless included in the present study to provide a complete treatment of the errors. Without loss of generality, the covariance matrix in this study is written as  $\Xi_{\text{meas}}=\Xi_{\text{pixel}} + \Xi_{\text{MS}} + \Xi_{\text{$\Delta$E}} = $ \text{diag}$\left\lbrace \sigma_u^2,\sigma_z^2,\sigma_{\phi}^2,\sigma_{\lambda}^2,\sigma_k^2 \right\rbrace$ with $\sigma_u=\sigma_z=\sigma_l$.

\subsection{Least-squares algorithm}\label{least-square}
\begin{figure*}[t]
\subfloat[][]
{\includegraphics[width=.50\textwidth]{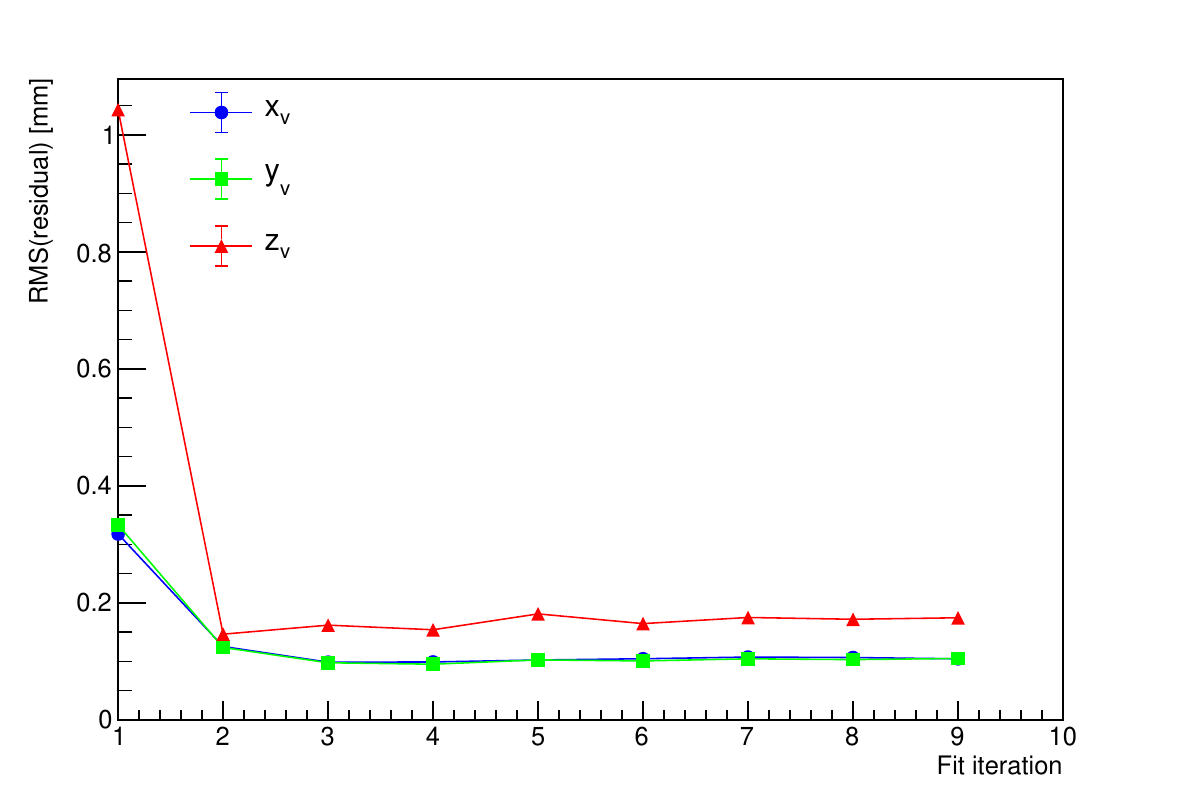}} 
\subfloat[][]
{\includegraphics[width=.50\textwidth]{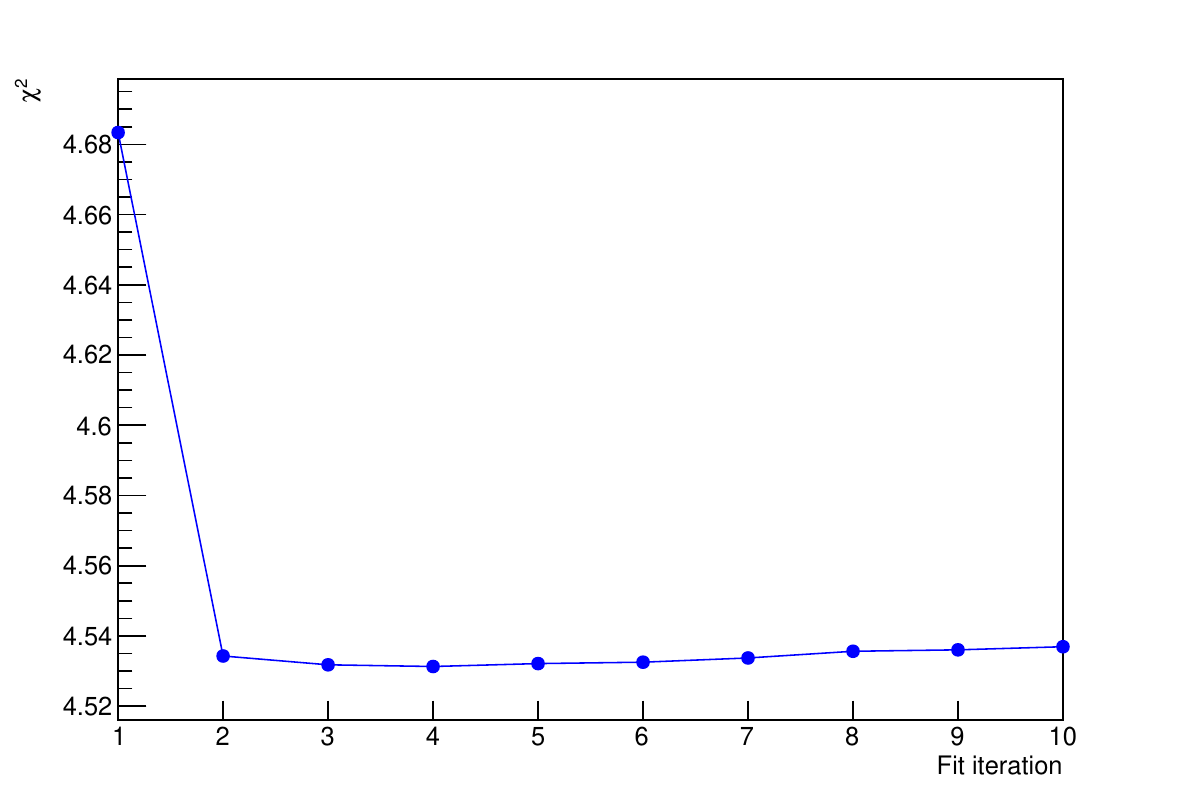}}
\caption{Typical RMS of the residuals $(x,y,z)_{meas} - (x,y,z)_{fit}$ at the RS (a) and the $\chi^2$ minimization (b) as a function of the fit iteration number.}\label{Res_Chi2}
\end{figure*}

In the vertex fitting, a map that links the track vertex parameters ($x,y,z,\phi^i,\lambda^i, k^i)_{\text{v}}$ to the parameters at the RS ($u^i,z^i,\phi^i,\lambda^i,k^i)_{\text{fit}}$ is needed, as shown in figure \ref{LocalRefFrame}. In what follows, $i =1,2,3$ for the ($e^-,e^+,e^+$) in a Mu3e decay. Starting from the analytical expression of a helical trajectory of a charged particle in a magnetic field, this map can be written as: $h_j=h(\textbf{v}, \textbf{t}, k_\text{v})_j$, where $\textbf{v} = (x,y,z)_{\text{v}}$, $\textbf{t}=(\phi,\lambda)_{\text{v}}$ and $h_j = (u^i,z^i,\phi^i,\lambda^i,k^i)_{\text{fit}}$ for $j=1,2,...,5$, see \ref{appendix1} for the analytic expressions of $h_j$. In first approximation, this function can be linearised near some initial guessed vertex parameters ($\textbf{v}_o, \textbf{t}_o, k_o$):

\begin{equation}\label{hmap}
\begin{aligned}
&h(\textbf{v}_o +\delta\textbf{v}, \textbf{t}_{i,o} +\delta\textbf{t}_i, k_{i,o} + \delta k_{i})_j \simeq\\ &h(\textbf{v}_o, \textbf{t}_{i,o}, k_{i,o})_j +  D_i\,\delta\textbf{v} +  E_i\,\delta\textbf{t}_i\, + F_i\,\delta k_i,
\end{aligned}
\end{equation}
where, for each track $i$, the matrices $D_i$, $E_i$ and $F_i$ have dimensions $5\times3$, $5\times2$ and $5\times1$, respectively, and are calculated as follows:

\begin{eqnarray}\label{hmapII}
\begin{array}{lll}
D^j_{n} = \frac{\partial h_j}{\partial \textbf{v}_n} \hspace{0.5cm} \textbf{v}_n=x_\text{v},y_\text{v},z_\text{v}\, \hspace{0.5cm} \text{for } n=1,2,3,\\
E^j_{m} = \frac{\partial h_j}{\partial \textbf{t}_m}\hspace{0.5cm} \textbf{t}_m=\phi_\text{v},\lambda_\text{v}\, \hspace{0.5cm} \text{for } m=1,2 ,\\
F^j =  \frac{\partial h_j}{\partial k_\text{v}} 
\end{array}
\end{eqnarray}

From the definitions in eq. \ref{hmap}, a quadratic cost function, $\chi^2$, is defined:
\begin{equation}\label{Chi2_II}
\begin{aligned}
\chi^2 &= \sum_i \Delta^T W_i \Delta\, ,\\
\Delta &= \delta q_i - D_i \delta\textbf{v} - E_i \delta\textbf{t}_i - F_i\delta k_i\, .\\
\end{aligned}
\end{equation}
where $W = \Xi^{-1}$ is the weight matrix (see for instance \cite{BILLOIR1985115}) and $\delta q$ is the residual at the RS:
\begin{equation}\label{Residual}
\delta q^i  = (u^i,z^i,\phi^i,\lambda^i,k^i)_{\text{meas}}-h(\textbf{v}_o, \textbf{t}^i_o,  k_o^i)\,.
\end{equation}
Equation \ref{Chi2_II} is the total normalized error, due to the approximation in eq. \ref{hmap}, expressed as a function of $\delta\textbf{v}$, $\delta\textbf{t}$ and $\delta k$. The accuracy of such an approximation can be seen in figure \ref{Res_Chi2}(a) where the RMS of the deviations $(x,y,z)_{\text{meas}} - (x,y,z)_{\text{fit}}$ at the RS are plotted versus fit iteration number. 

The corrections to the initial guessed vertex parameters can be found by minimizing the cost function, i.e., by solving the following system of equations:
\begin{equation}\label{chi2_system2}
\left\lbrace
\begin{aligned}
 &\sum_i A_i^T\delta\textbf{v} + \sum_i B^T_i \delta\textbf{t}_i + \sum_i H^T_i\delta k_i = \sum_i T_i,\\
 &B_i\delta\textbf{v} + C_i\delta\textbf{t}_i + G_i\delta k_i = U_i,\\
 &H_i\delta\textbf{v} + G_i^T\delta\textbf{t}_i + L_i\delta k_i =Z_i.
\end{aligned}
\right.
\end{equation}
where
\begin{equation}\label{EqMatricesII}
\begin{aligned}
&A_i^T = D_i^T W_iD_i\, ,\hspace{0.5cm} B_i = E_i^T W_iD_i\,\hspace{0.5cm} C_i=E_i^TW_iE_i\, ,\\
&G_i = E_i^T W_iF_i\, , \hspace{0.5cm} H_i = F^TW_iD_i\, ,\hspace{0.5cm}  L_i=F_i^TW_iF_i\, ,\\
&T_i = D_i^TW_i\delta q_i\, ,\hspace{0.5cm}  U_i=E_i^TW_i\delta q_i\,  \hspace{0.5cm} Z_i = F_i^TW_i\delta q_i .
\end{aligned}
\end{equation}

The solutions of the system described in eq. \ref{chi2_system2} are:
\begin{equation}\label{EqBilloirFit2}
\begin{aligned}
 \delta \textbf{v} &= \left[ \left(\sum_i A^T - \sum_i B^T_i C^{-1}_i B_i \right) - M N_1N_3\right] ^{-1}\\
 &\times \left(\sum_i T_i - \sum_i B_i^T C^{-1}_i U_i - M N_1N_2 \right), \\
 \delta k_i &= N_1\left(N_2 - N_3\delta \textbf{v}\right),\\
 \delta \textbf{t}_i &= C_i^{-1}\left( U_i - B_i \delta\textbf{v} - G_i \delta k_i\right),
\end{aligned}
\end{equation}
where:
\begin{equation}\label{EqBilloirFit22}
\begin{aligned}
  M   &= \left( \sum H_i^T - \sum_i B^T_i C^{-1}_i G_i \right)\, ,\\
  N_1 &= \left( L_i - G^T_i C^{-1}_i G_i \right)^{-1}  \, , \\
  N_2 &= \left( Z_i - G^T_i C^{-1}_i U_i \right)\, \\
  N_3 &= \left( H_i - G^T_i C^{-1}_i B_i \right) \,.
\end{aligned}
\end{equation}
From eq. \ref{EqBilloirFit2} the covariance matrices for the track parameters at the vertex can be calculated: 

\begin{equation}\label{CovEqs}
\begin{aligned}
 \text{Cov}(\textbf{v}) &= \left[ \left(\sum_i A^T - \sum_i B^T_i C^{-1}_i B_i \right) - M N_1N_3\right] ^{-1},\\
 \text{Cov}(k_i) &= L_i^{-1} + \left( N_1N_3\right) \text{Cov}(\textbf{v})\,\left( N_1N_3 \right)^T,\\
 \text{Cov}(\textbf{t}_{i}) = &C_i^{-1} + \left(C_i^{-1}B_i\right)\text{Cov}(\textbf{v}) \left(C_i^{-1}B_i\right)^T +\\
 &\left(C_i^{-1}G_i\right) \text{Cov}(k)\left(C_i^{-1}G_i\right)^T.
\end{aligned}
\end{equation}

\subsection{Algorithm testing}\label{Test-results}
Geant4 based Monte-Carlo (MC) simulations of Mu3e decays have been carried out for testing the vertex fitting described in section \ref{least-square}. The procedure followed by the test was:
\begin{enumerate}
\item Hit parameters at the RS, $(u^i,z^i,\phi^i,\lambda^i,k^i)_{MC}$, were obtained from MC trajectories. 
\item The parameters were smeared by adding a random offset according to the errors in $\Xi_{\text{meas}}$.
\item The initial parameters $\textbf{v}_o=x_o,y_o,z_o$ were obtained as the average coordinates of the tracks intersection points or points of closest approach. Since two tracks can have up to two intersections, the one that is encountered first during the propagation from the RS to the target is retained in the calculation of the average value. The vectors $\textbf{t}_o$ were extracted at the point of closest approach of the trajectories to $\textbf{v}_o$ whilst $k_o$ was directly obtained from the track reconstruction carried out before the vertex fitting.  
\item From equations \ref{hmap} and \ref{hmapII}, the residuals at the RS were calculated and thus the corrections $\delta \textbf{v}$, $\delta k_i$ and $\delta \textbf{t}_i$ in eq. \ref{EqBilloirFit2}. Only a few iterations were required to minimize the $\chi^2$ in eq. \ref{Chi2_II}, as it can be seen in figure \ref{Res_Chi2}(b).
\end{enumerate}

\begin{figure}[h!]
\centering
\subfloat[][]
{\includegraphics[width=.5\textwidth]{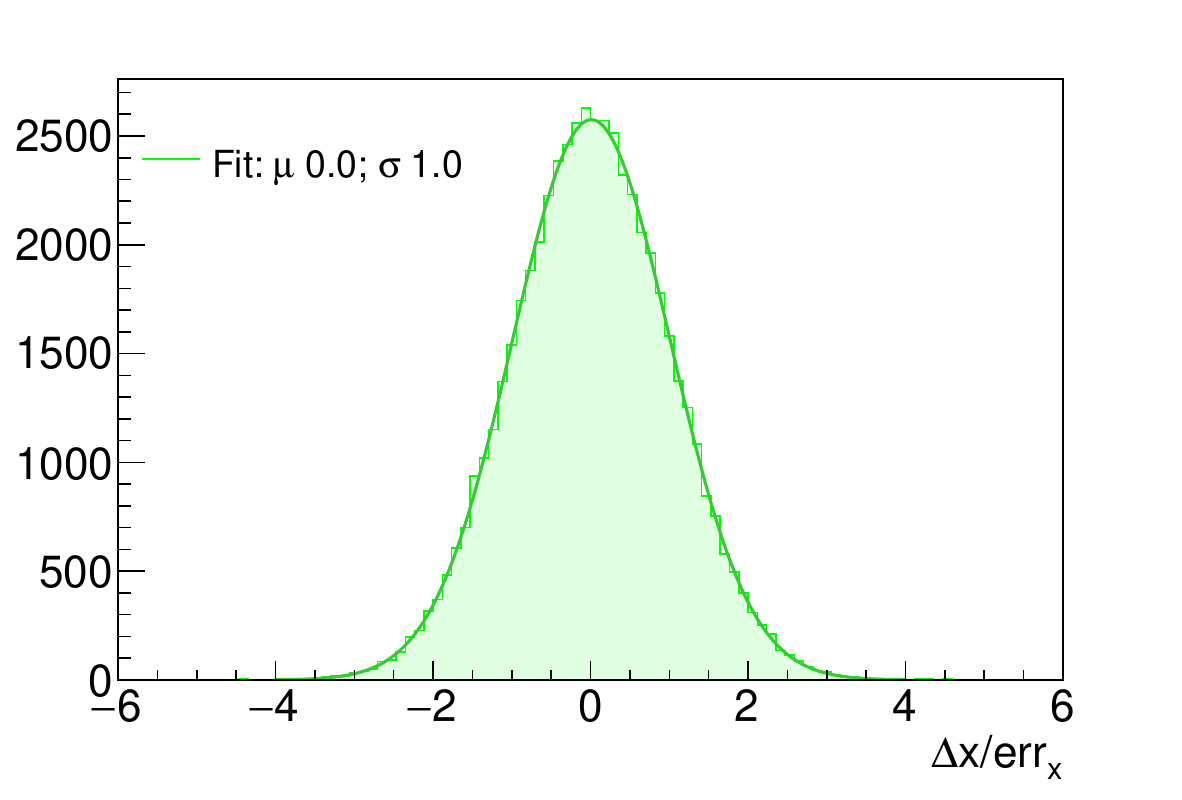}} 
\subfloat[][]
{\includegraphics[width=.5\textwidth]{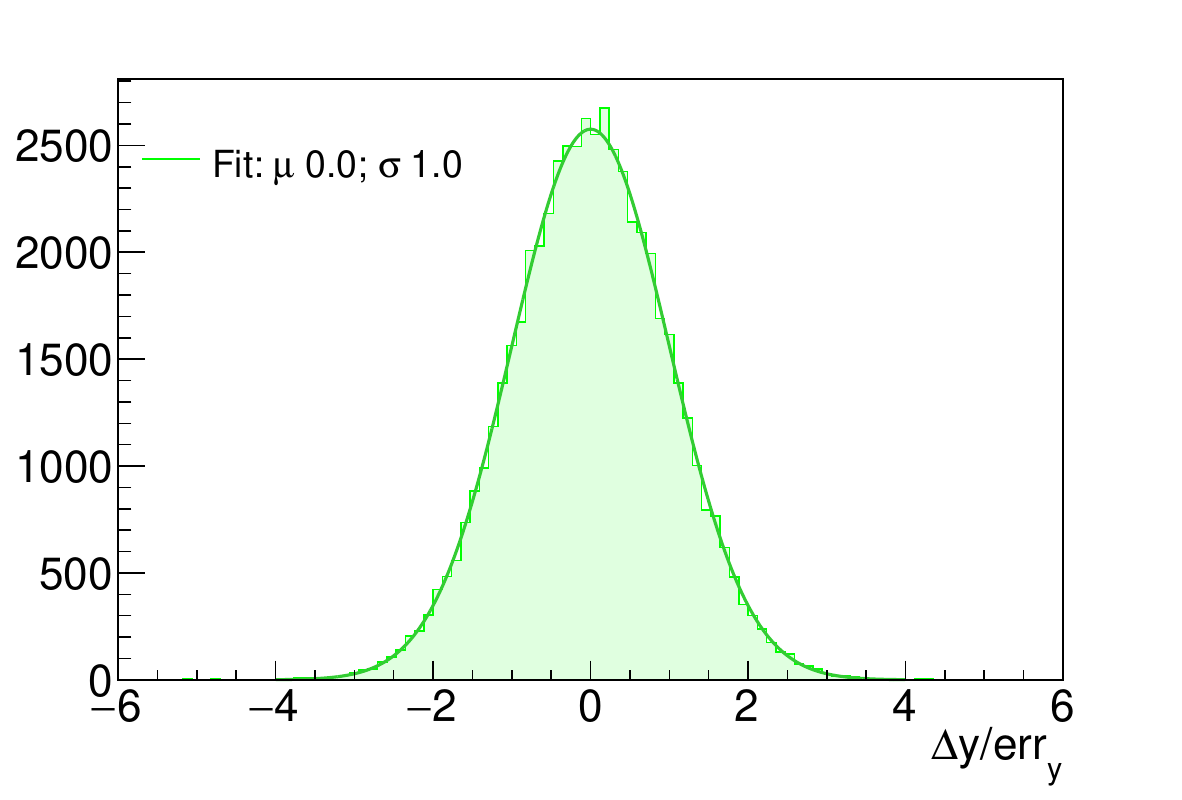}} \\
\subfloat[][]
{\includegraphics[width=.5\textwidth]{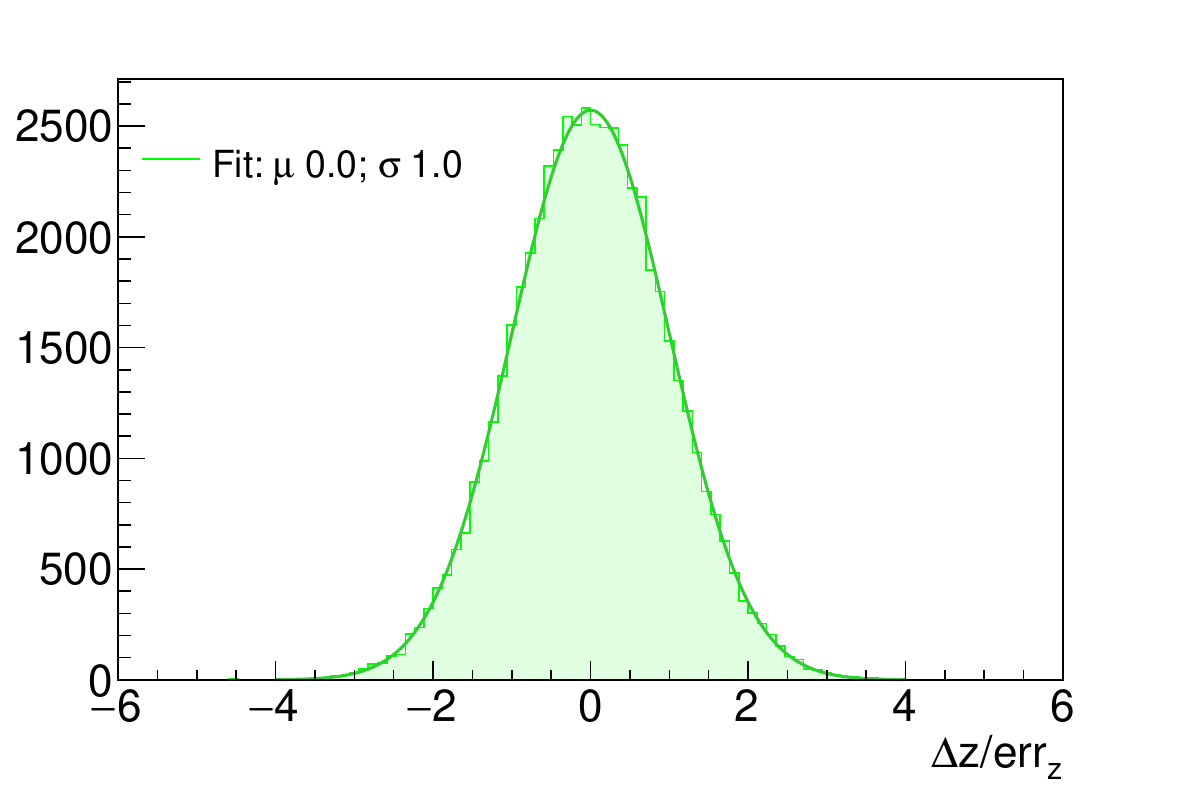}} 
\subfloat[][]
{\includegraphics[width=.5\textwidth]{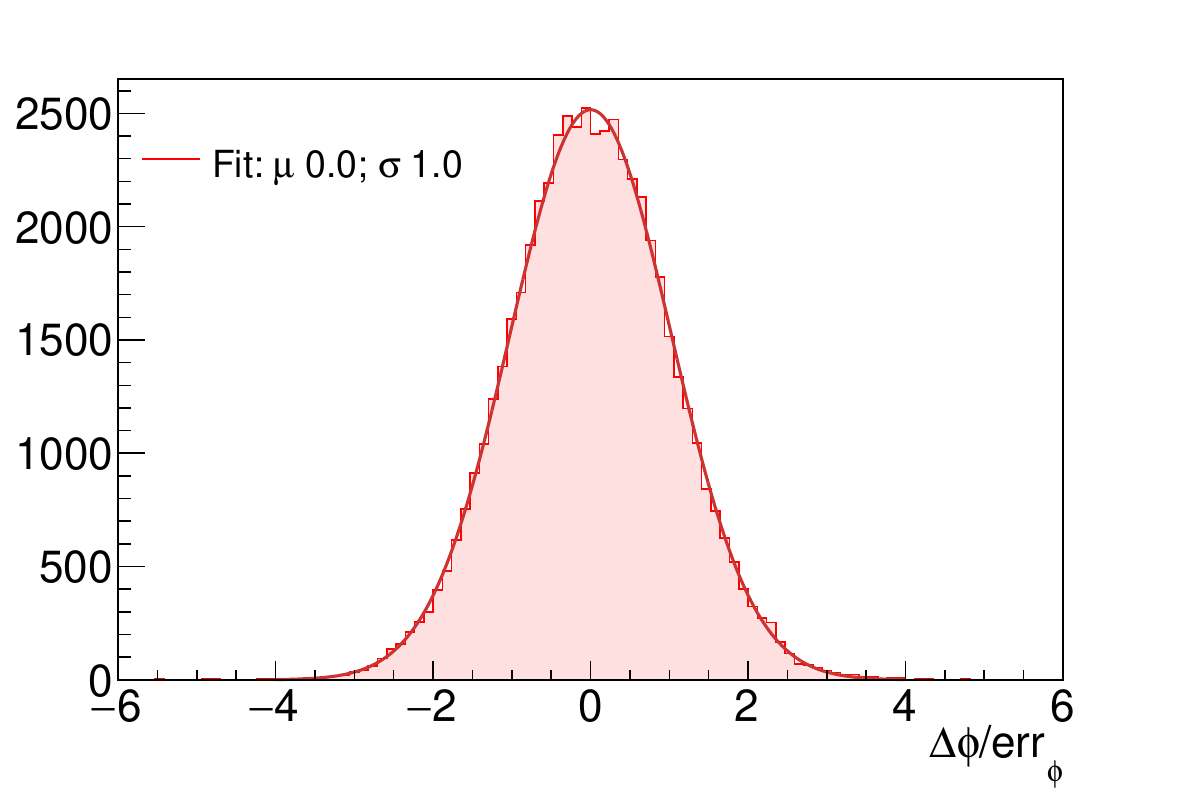}}\\
\subfloat[][]
{\includegraphics[width=.5\textwidth]{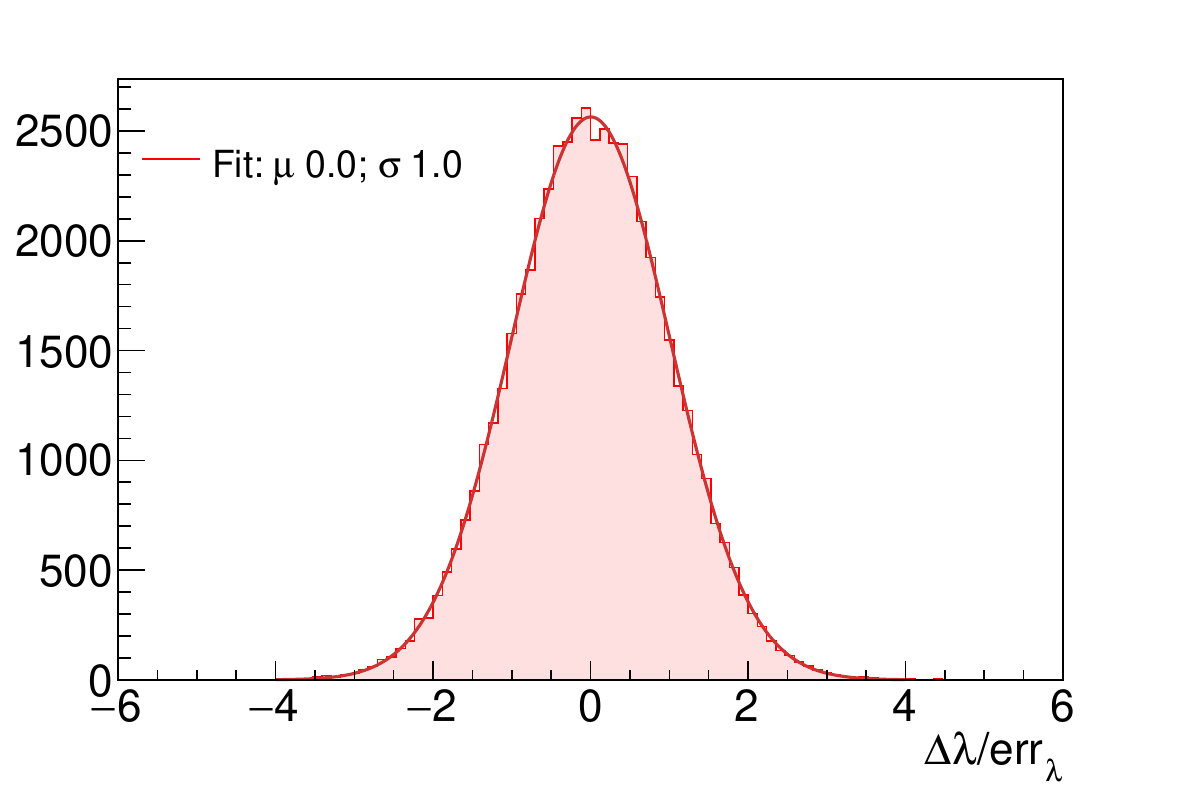}}
\subfloat[][]
{\includegraphics[width=.5\textwidth]{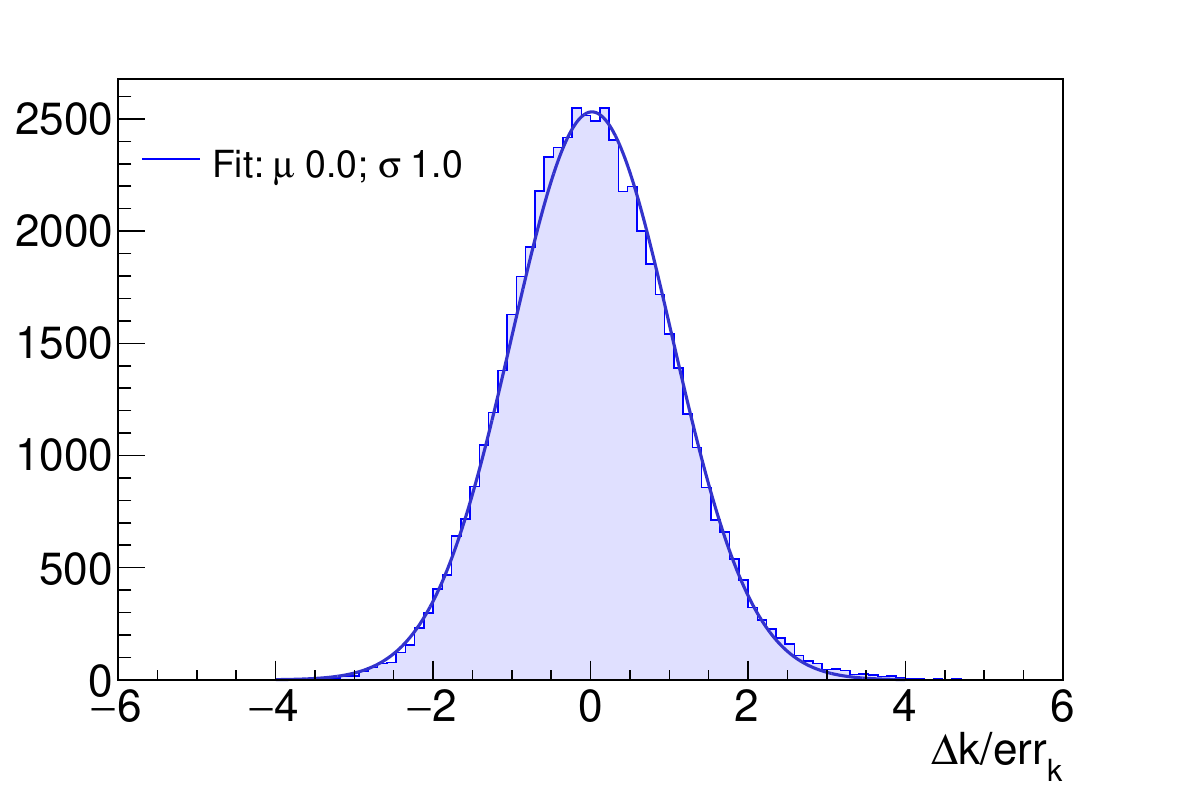}}
\caption{Pull distributions (a-f) of the track vertex parameters: ($x,y,z,\phi,\lambda,k)_{\text{v}}$, as calculated in the test procedure.}\label{Pulls}
\end{figure}

A key test of the algorithm consists in plotting the pull distributions of the track vertex parameters obtained from the fit. The pull of a variable $X$ with expected value $\mu_X$ and standard error $\sigma_X$ is:
\begin{equation}\label{Eqpull}
P = \frac{\left( X - \mu_X\right) }{\sigma_X}.
\end{equation}
If the equations \ref{EqBilloirFit2} and \ref{CovEqs} are correct, the derived pulls should follow a normal distribution. Figure \ref{Pulls} displays the normal distributions for the pulls of all the track parameters at the vertex, as calculated in this test.

\section{A comparative study of error sources}\label{Mu3eCaseStudy}
\begin{figure}[t!]
\centering
\includegraphics[scale=0.5]{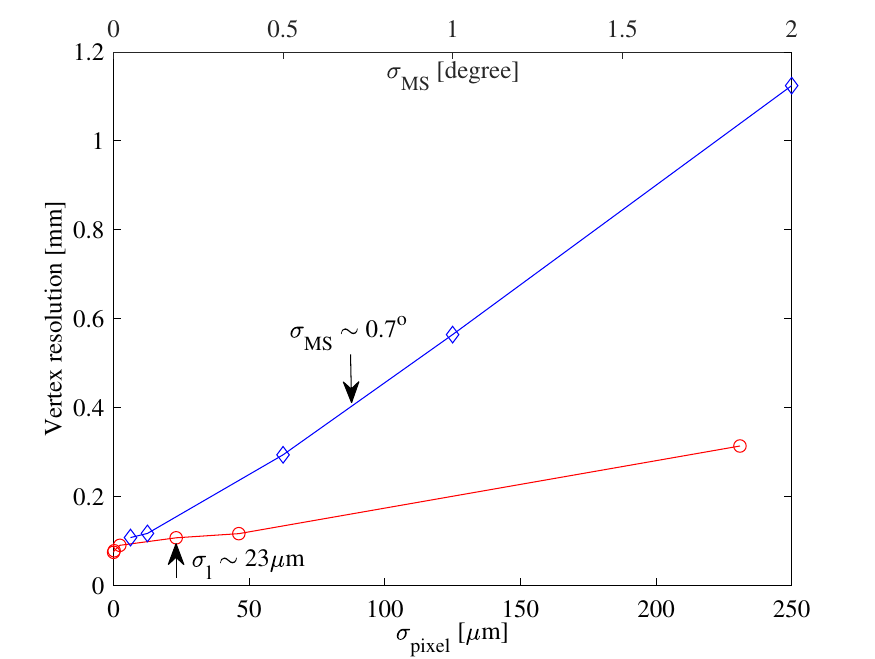}
\caption{Blue diamonds, vertex resolution as a function of the MS for $\sigma_l=l/\sqrt(12)/1000$. Red circles,vertex resolution as a function of the $\sigma_l$ for MS $=0.0001^{\circ}$. The two arrows indicate the values of $\sigma_{MS}$ and $\sigma_l$ according to the Mu3e experimental design.}\label{VertexRes_Vs_MS_PX}
\end{figure}

\begin{figure*}[htpb]
\centering
\subfloat[][]
{\includegraphics[width=.50\textwidth]{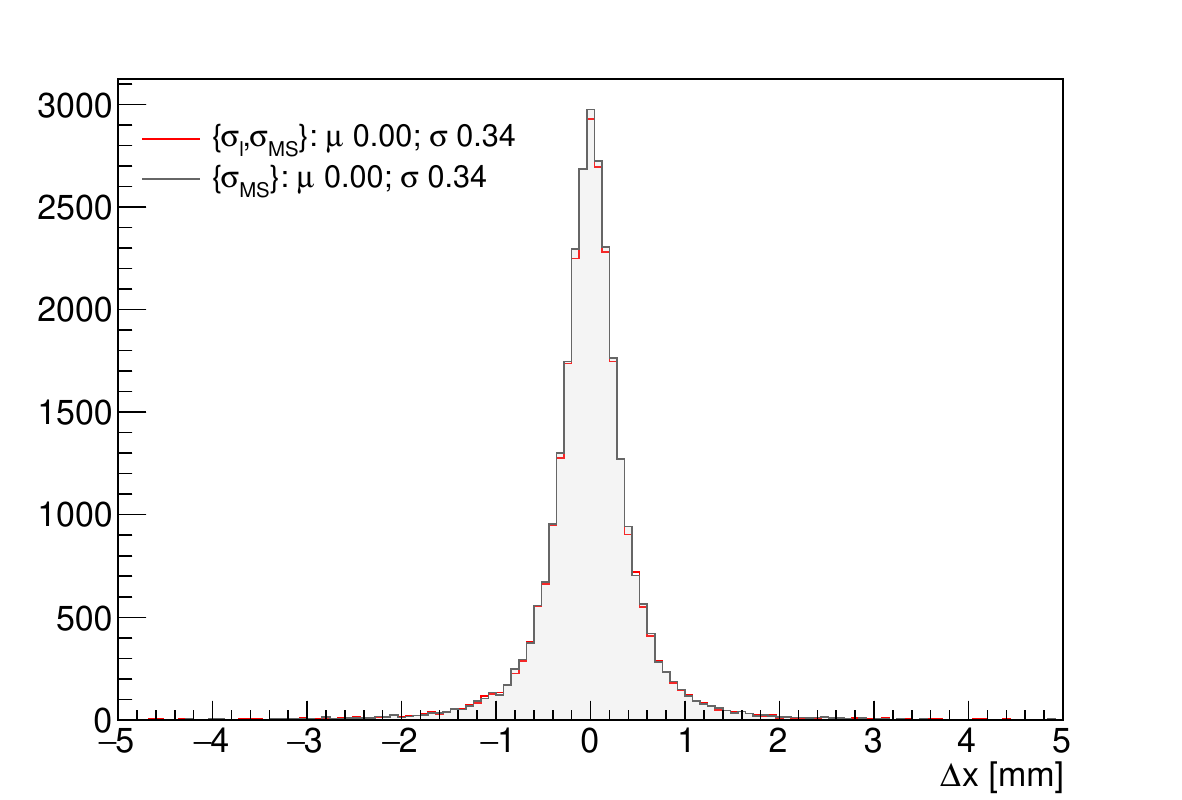}} 
\subfloat[][]
{\includegraphics[width=.50\textwidth]{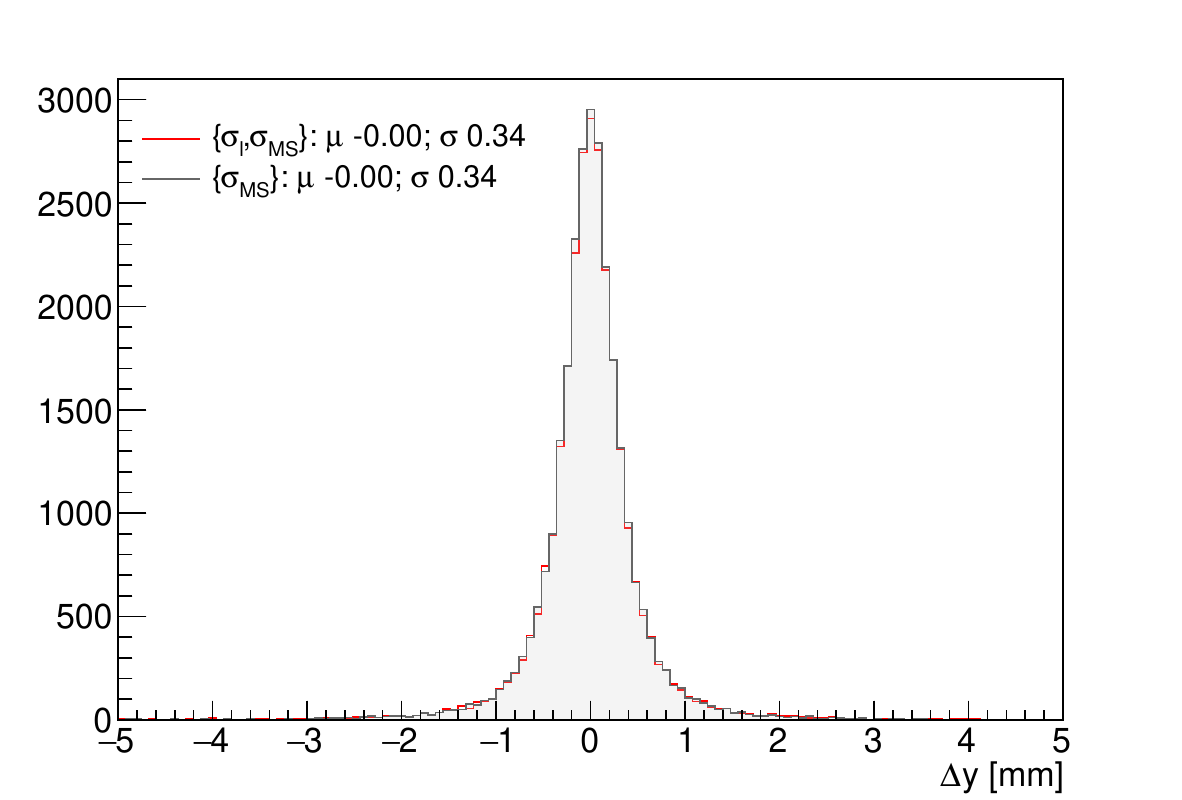}} \\
\subfloat[][]
{\includegraphics[width=.50\textwidth]{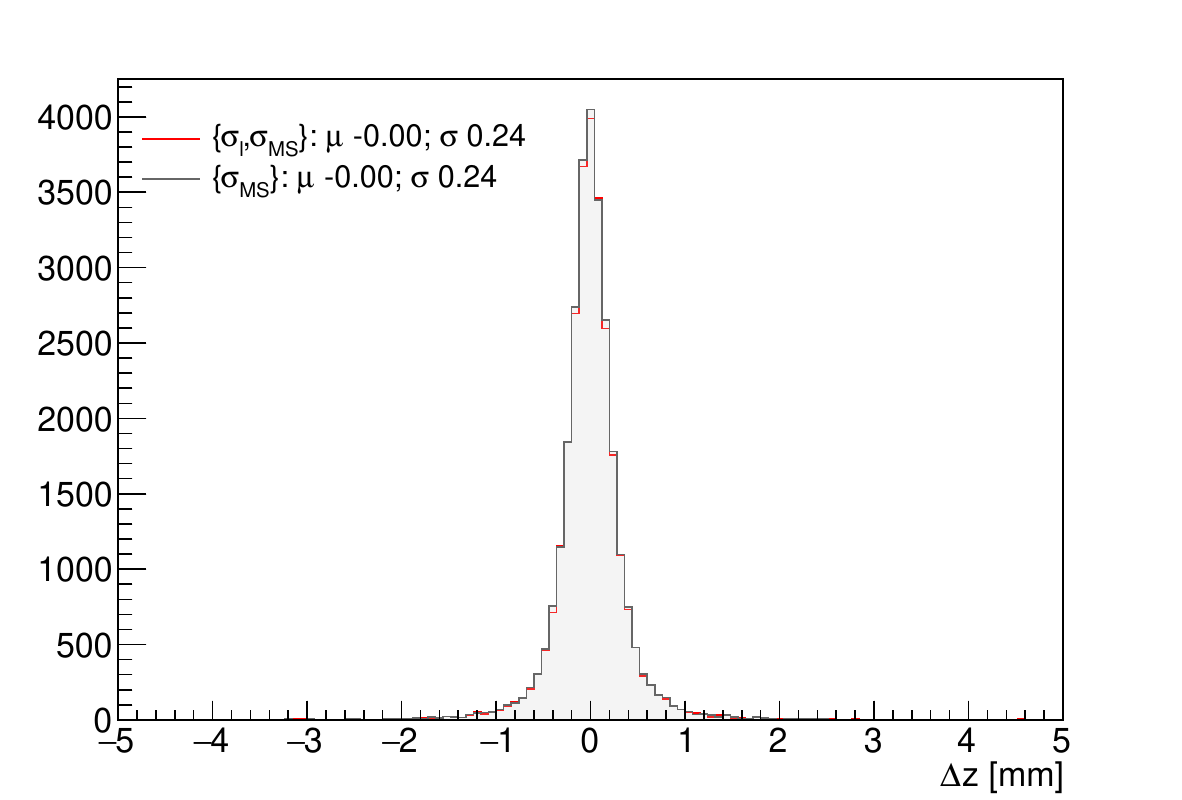}} 
\subfloat[][]
{\includegraphics[width=.50\textwidth]{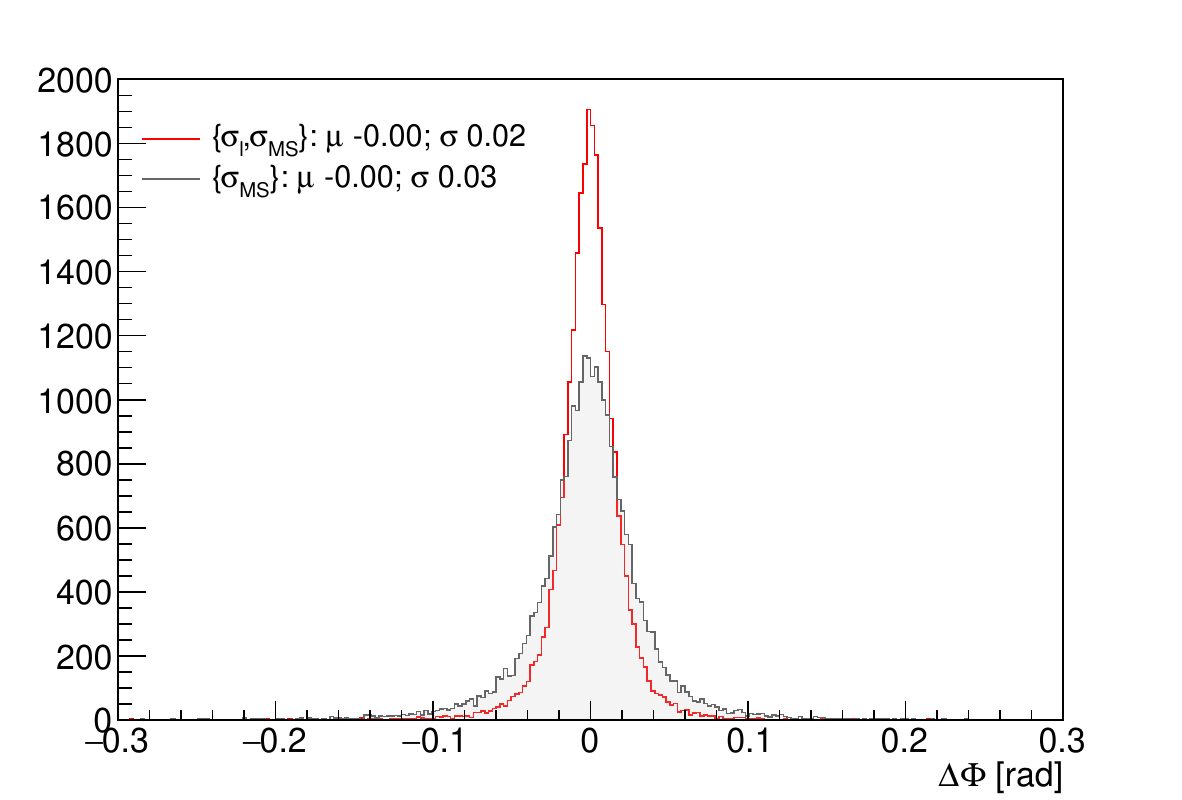}}\\
\subfloat[][]
{\includegraphics[width=.50\textwidth]{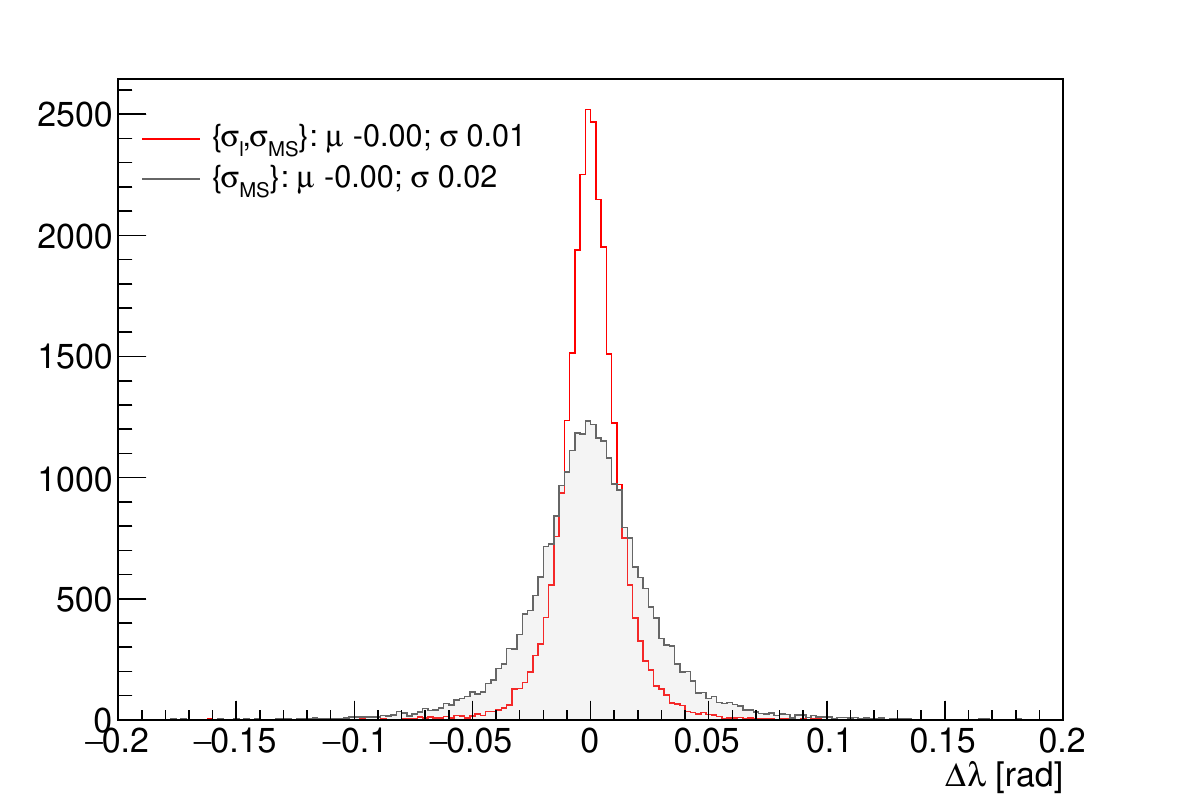}}
\subfloat[][]
{\includegraphics[width=.50\textwidth]{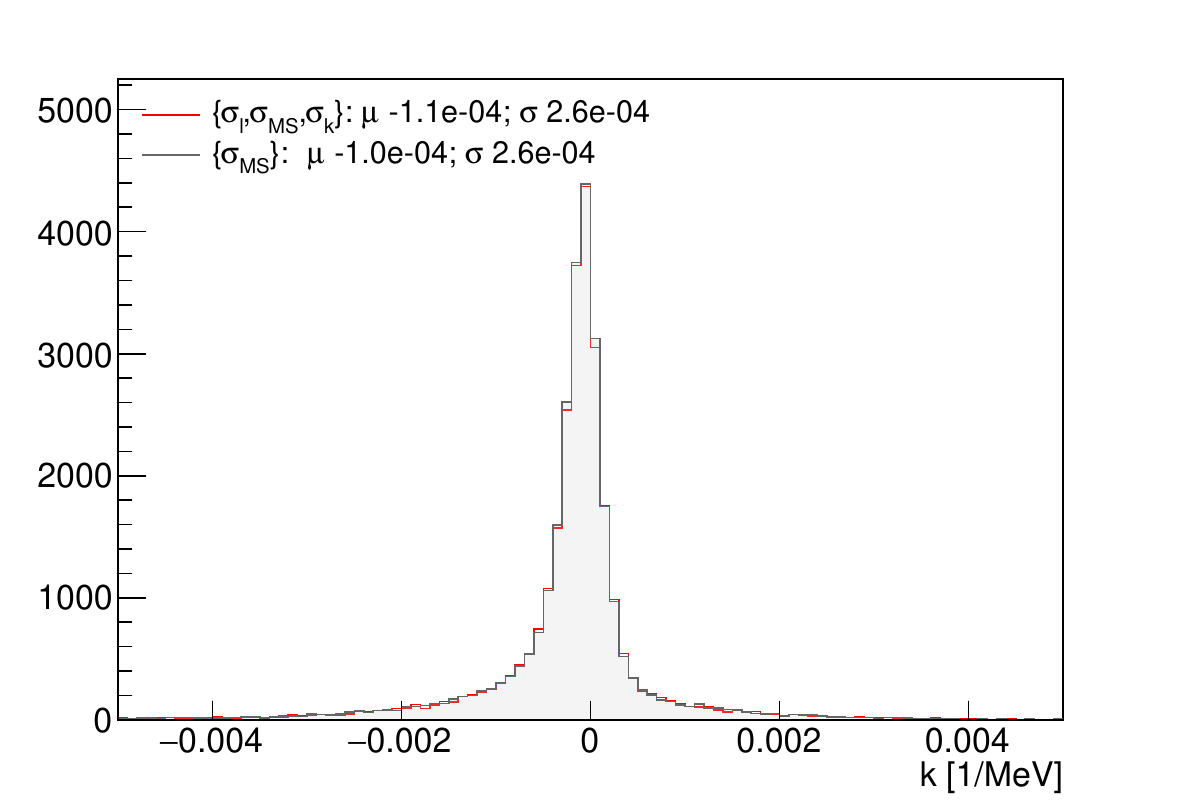}}
\caption{Deviations between the fit and MC track parameters at the vertex. The legends show the mean and standard deviation obtained from a Gaussian fit in scenarios (A) [full gray] and (B or C) [empty red], respectively.}\label{ComparisonMethod1}
\end{figure*}

To investigate the effects of pixel resolution and MS on the vertex resolution, various random offsets have been applied to the MC test trajectories simulated in section \ref{Test-results}. This allowed us to target resolution scenarios dominated by MS, i.e., $\sigma_{MS}\gg\sigma_l$, or the intrinsic detector spatial resolution for $\sigma_l\gg\sigma_{MS}$.  Figure \ref{VertexRes_Vs_MS_PX}, shows the contribution of pixel resolution and MS to vertex resolution (defined as $\sqrt{\left(\sigma_x^2 + \sigma_y^2 + \sigma_z^2 \right) }$, see panels (a-c) in figure \ref{ComparisonMethod1}). It can be seen that MS dominates over pixel resolution for a wide range of $\sigma_l$ and $\sigma_{MS}$ values, and it is a factor 4 times higher for the Mu3e pixel detector, i.e.,  $\sigma_{MS} \sim 0.7^{\circ}$ and $\sigma_l \sim 23$\SI{}{\micro\meter}. Despite being smaller in magnitude, the contribution of pixel resolution to the final fit accuracy is not negligible. This can be appreciated by studying three inclusive scenarios: (A) MS only; (B) MS and pixel resolution are both included; (C) all sources of errors (MS, pixel resolution and energy losses) are considered. These three scenarios are summarized in table \ref{Tab1}.

It is important to note that the kernel of $\Xi_{\text{meas}}$ grows larger going from scenario (A) to (C) along with the dimensionality of the problem, see for example \citep{BILLOIR1992139}. For instance, if the energy loss and pixel resolution at the RS are ignored, MS remains the only source of uncertainty against which all measurements are fixed but $(\phi_i,\lambda_i)_{\text{meas}}$, i.e., $\Xi_{\text{meas}}=\text{diag} \left\lbrace \sigma_{\phi}^2,\sigma_{\lambda}^2\right\rbrace$. These 6 angles can be fitted with 3 vertex variables $(x,y,z)_{\text{v}}$ thus simplifying eq. \ref{EqBilloirFit2}, see e.g., \cite{schenk2013vertex}. 

Panels (a-e) in figure \ref{ComparisonMethod1} show the the deviations between MC vertex parameters of simulated Mu3e decays and those obtained from vertex fitting in scenarios (A) and (B), respectively. From figure \ref{ComparisonMethod1}(a-c), it can be seen that the fit accuracy for the determinations of $(x,y,z)_{\text{v}}$ does not improve when the pixel resolution is included in $\Xi_{\text{meas}}$. However, a significant improvement is found on the determinations of $(\phi_i,\lambda_i)_{\text{v}}$, as shown in figure \ref{ComparisonMethod1}(d,e). The results for scenario (C) are statistically the same as for scenario (B). The former being the only case in which the fit attempts to optimize the track parameter $k$, as detailed in table \ref{Tab1}. As expected, having neglected energy losses at the RS, track curvatures do not vary significantly throughout the vertex fitting, as it can be seen in figure \ref{ComparisonMethod1}(f). 

\begin{figure*}[h!]
\centering
\subfloat[][]
{\includegraphics[width=.5\textwidth]{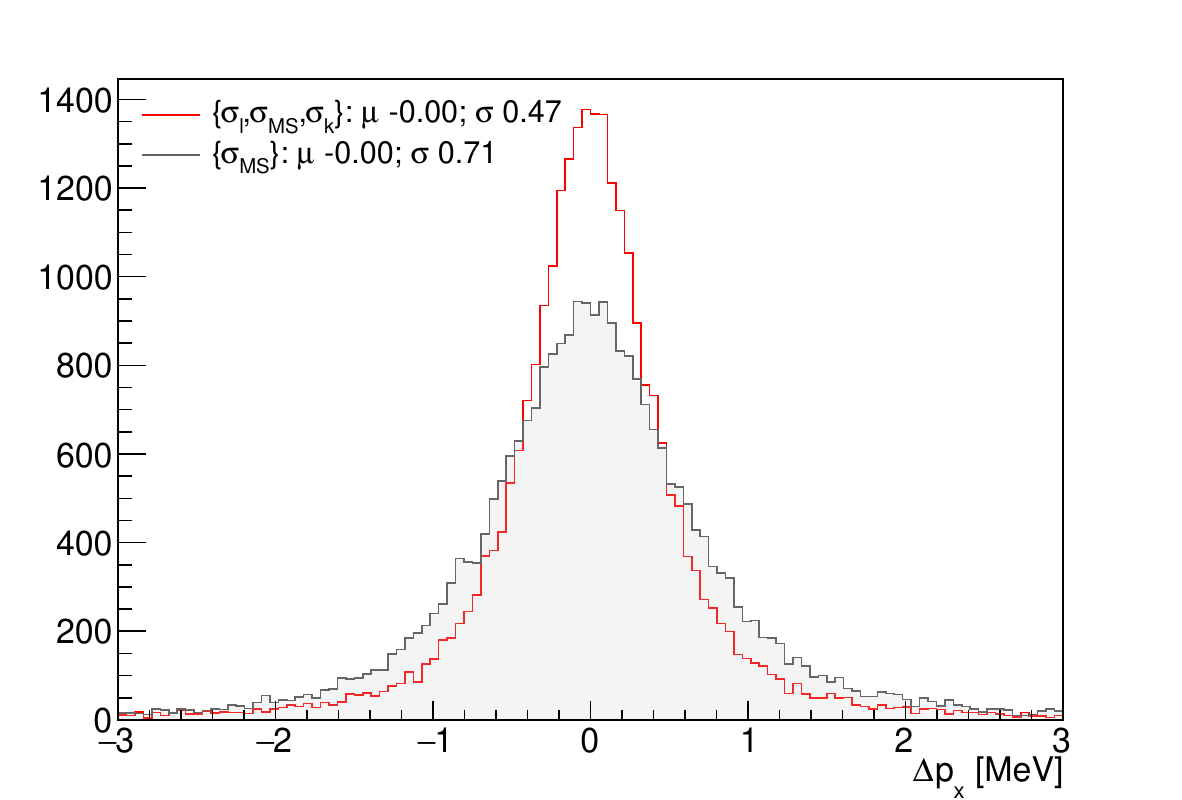}} 
\subfloat[][]
{\includegraphics[width=.5\textwidth]{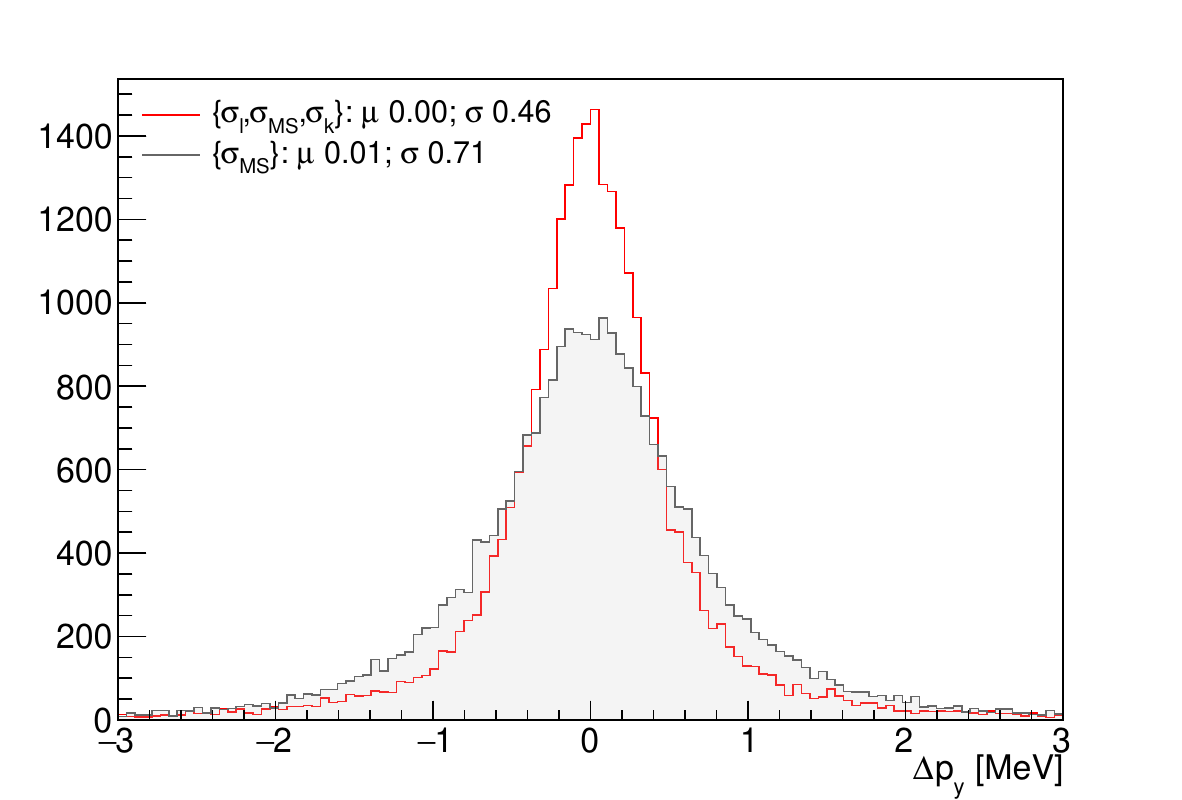}}\\
\subfloat[][]
{\includegraphics[width=.5\textwidth]{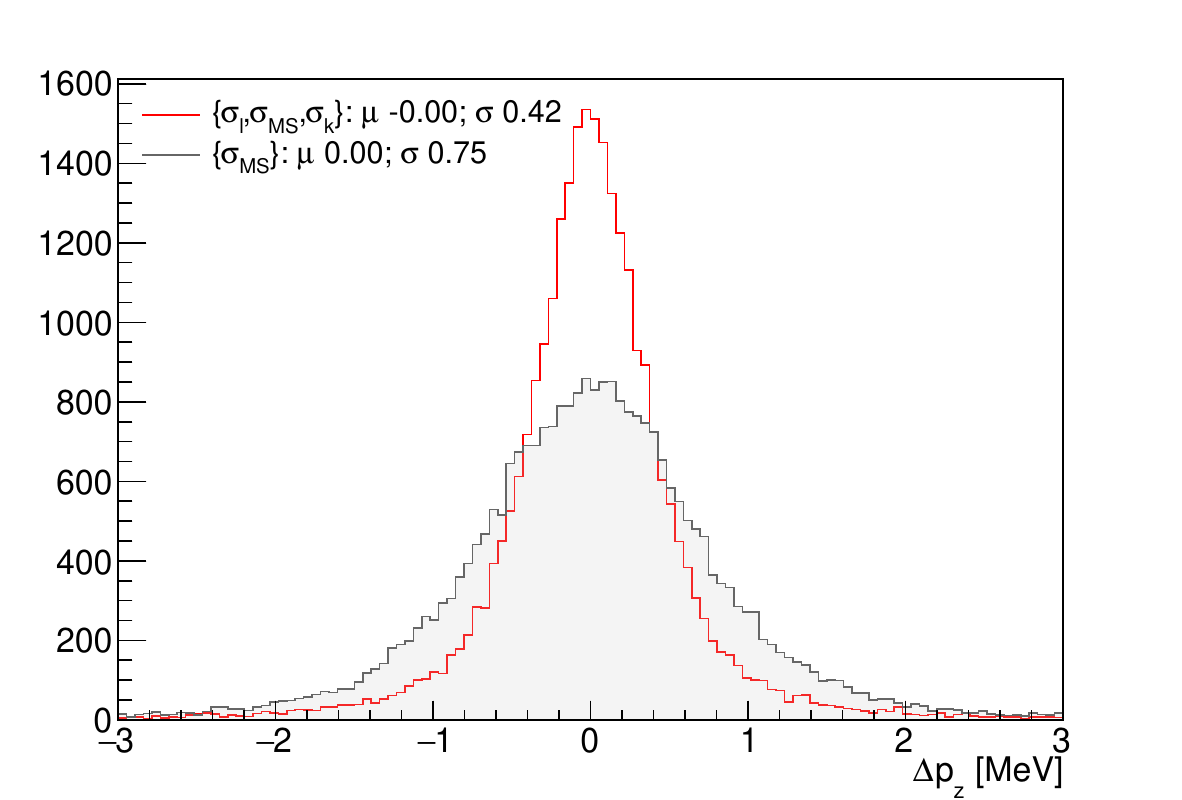}}
\subfloat[][]
{\includegraphics[width=.5\textwidth]{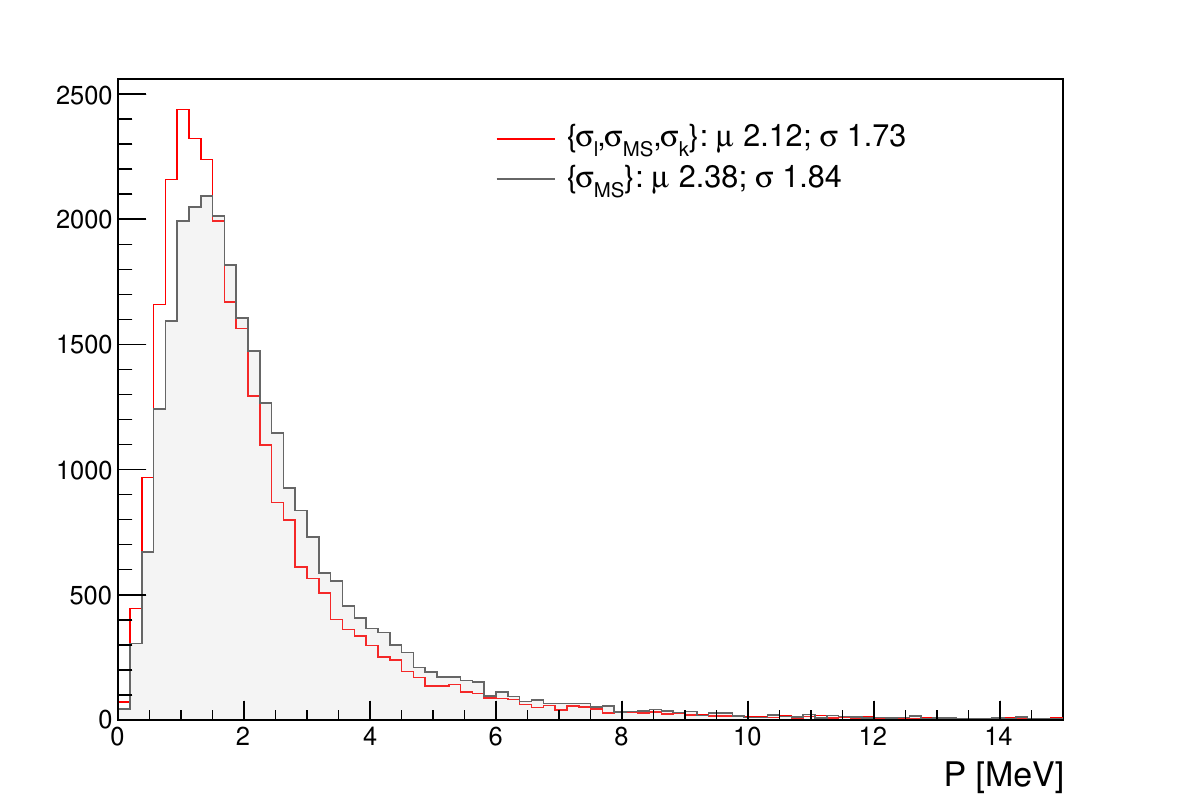}}
\caption{Deviations, at the vertex, between the fit and MC momentum coordinates $p_x$, $p_y$ and $p_z$ panels (a-c) and total momentum $P$ in panel (d) for scenario (A) [full gray] and (C) [empty red], respectively. In (a,b,c), the legends show the mean and standard deviation obtained from a Gaussian fit whilst the legend in (d) shows the average and standard deviation of the distributions.}\label{ComparisonMethod2}
\end{figure*}

The improved fit accuracy of case (C) with respect to scenario (A) for the coordinates of the momentum vector $p_x$, $p_y$ and $p_z$ is shown in figure \ref{ComparisonMethod2}(a-c). This improvement reflects also onto the determination of the average total momentum which is $\sim$10$\%$ smaller in scenario (C) than the corresponding average in scenario (A), and thus closer to real MC value (in the hypothesis of decaying muons at rest), see figure \ref{ComparisonMethod2}(d).

\begin{figure}[h!]
\centering
\includegraphics[scale =0.5]{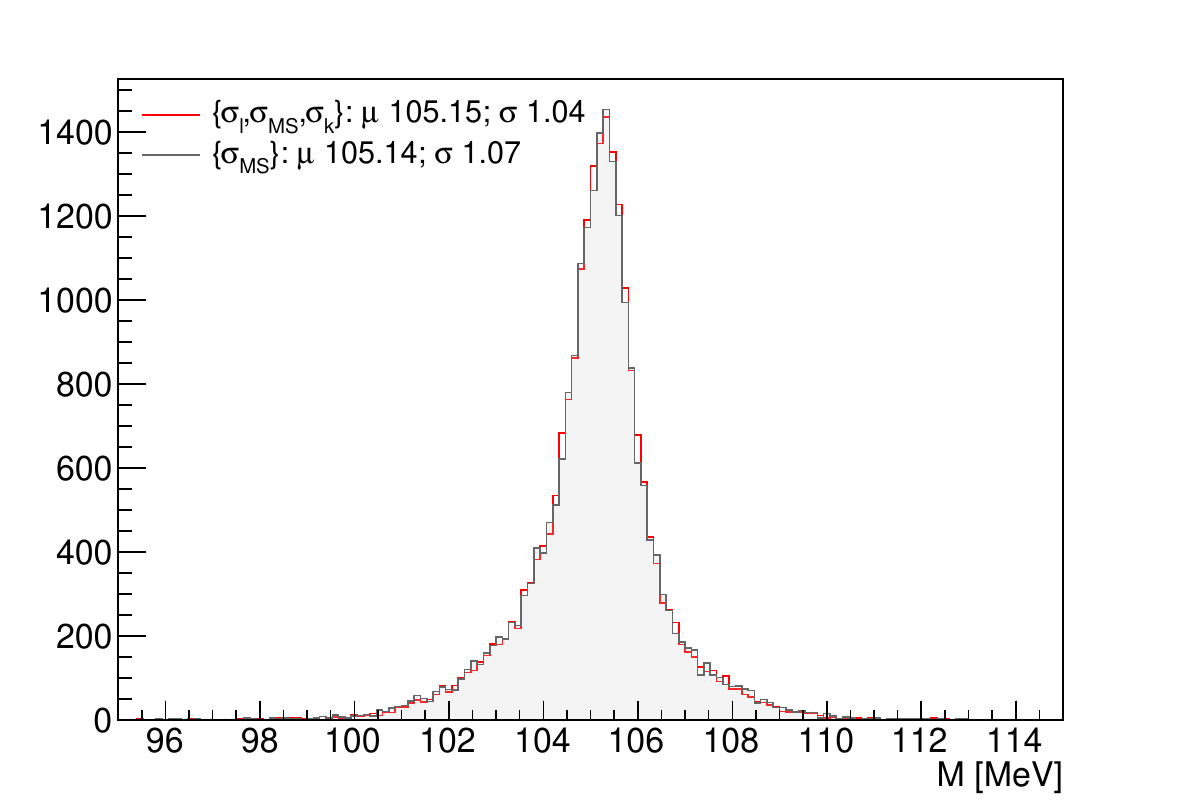}
\caption{Reconstructed invariant mass of Mu3e decays, scenario (A) [full gray] and (C) [empty red], for $\chi^2\leq15$ and $P\leq4$\,MeV/c.}\label{ComparisonMethod3}
\end{figure}

The reconstructed invariant mass of simulated Mu3e decays in scenarios (A) and (C) is shown in figure \ref{ComparisonMethod3}. As expected, no significant difference is seen in the two fit scenarios. In fact, the magnitude of the invariant mass is dominated by the muon rest mass over which the fit accuracy has little leverage.

\begin{table*}[t]
\centering
\caption{Scenarios A,B,C in the comparative study of the present vertex fitting algorithm.}\label{Tab1}
\footnotesize
\begin{tabular}{@{}ccccc}
Scenario &$\Xi_{\text{meas}}$&measurements& fit parameters&errors\\
\hline
A& $\text{diag} \left\lbrace \sigma_{\phi}^2,\sigma_{\lambda}^2\right\rbrace$ &$(\phi,\lambda)_{\text{meas}}$ &$(x,y,z)_{\text{v}}$ & MS\\
B& $\text{diag} \left\lbrace \sigma_u^2,\sigma_z^2,\sigma_{\phi}^2,\sigma_{\lambda}^2\right\rbrace$ &$(u,z,\phi,\lambda)_{\text{meas}}$ &$(x,y,z,\phi,\lambda)_{\text{v}}$ & MS, pixel\\
C& $\text{diag} \left\lbrace \sigma_u^2,\sigma_z^2,\sigma_{\phi}^2,\sigma_{\lambda}^2,\sigma_k^2\right\rbrace$ &$(u,z,\phi,\lambda,k)_{\text{meas}}$ &$(x,y,z,\phi,\lambda,k)_{\text{v}}$ & MS, pixel, $\Delta E$\\\\
\end{tabular}\\
\end{table*}

\section{Conclusion}
In this work, a fitting algorithm appositely designed for the reconstruction of decay vertices in light pixel detectors was developed and described. The study focused on the Mu3e low-material budget pixel detector, exploring the relative impact of multiple scattering, pixel resolution and energy losses on the vertex resolution. The results demonstrate a significant improvement in the fit accuracy when accounting for the resolution of the pixel sensors. This should encourage a rigorous consideration of detector intrinsic spatial resolution in the development of future reconstruction algorithms concerning precise particle physics measurements at low energy.

\section{Acknowledgements}
This work was supported by the STFC Consolidated Grant PPR10542. I wish to thank the members of the Mu3e Software and Analysis group for providing the simulation and track reconstruction software used in this study. Special thanks go Joel Goldstein, Niklaus Berger and Gavin Hesketh for their detailed and insightful discussions about this work and Naik Paras and Andre Sch\"{o}ning for their careful reading and valuable comments.

\appendix

\section{Propagation of track parameters}\label{appendix1}
The analytical expression of $h(\textbf{v}, \textbf{t}, k)$ is derived by starting from the propagation of a track in the transverse plane and then along the beam direction.

\subsubsection*{Propagation in the transverse plane}
In the transverse plane, trajectories are helices with symmetry axis $\hat{z}$ and transverse radius $R_{\perp}$ which sign is given by the charge $q$ of the particle, see eq. \ref{initialSystem}. We write $R_{\perp}=R$\,\text{cos($\lambda$)} such as:
\begin{equation}\label{Rsign}
R :=\left\lbrace
\begin{aligned}
 &+ \frac{p}{|q|B} \hspace{0.3cm} \text{if q } >0 \text{ c.c.w rotation}\,\\
 &- \frac{p}{|q|B} \hspace{0.3cm} \text{if q } <0 \text{ c.w rotation}\, .
\end{aligned}
\right.
\end{equation}
In the x-y plane, a helix is a circumference with center in $(x_c, y_c)$ and radius $R_{\perp}$. It is not difficult to prove that:

\begin{equation}\label{EqCenterPosition}
\begin{aligned}
&x_c = x_{\text{v}} - R_{\perp}\text{sin}(\phi_{\text{v}})\, ,\\
&y_c = y_{\text{v}} + R_{\perp}\text{cos}(\phi_{\text{v}})\, ,\\
\end{aligned}
\end{equation}

The transport equations in the transverse plane is obtained by calculating the coordinates $(x_q,y_q)$ of the intersection point between the track originating from $(x_\text{v},y_\text{v})$ and the detector ladder. In the x-y plane, the ladder profile is a line characterised by the parameters $y_o$ and $m$, see figure \ref{Intersection}:

\begin{figure}[h!]
\begin{center}
\includegraphics[scale=0.5]{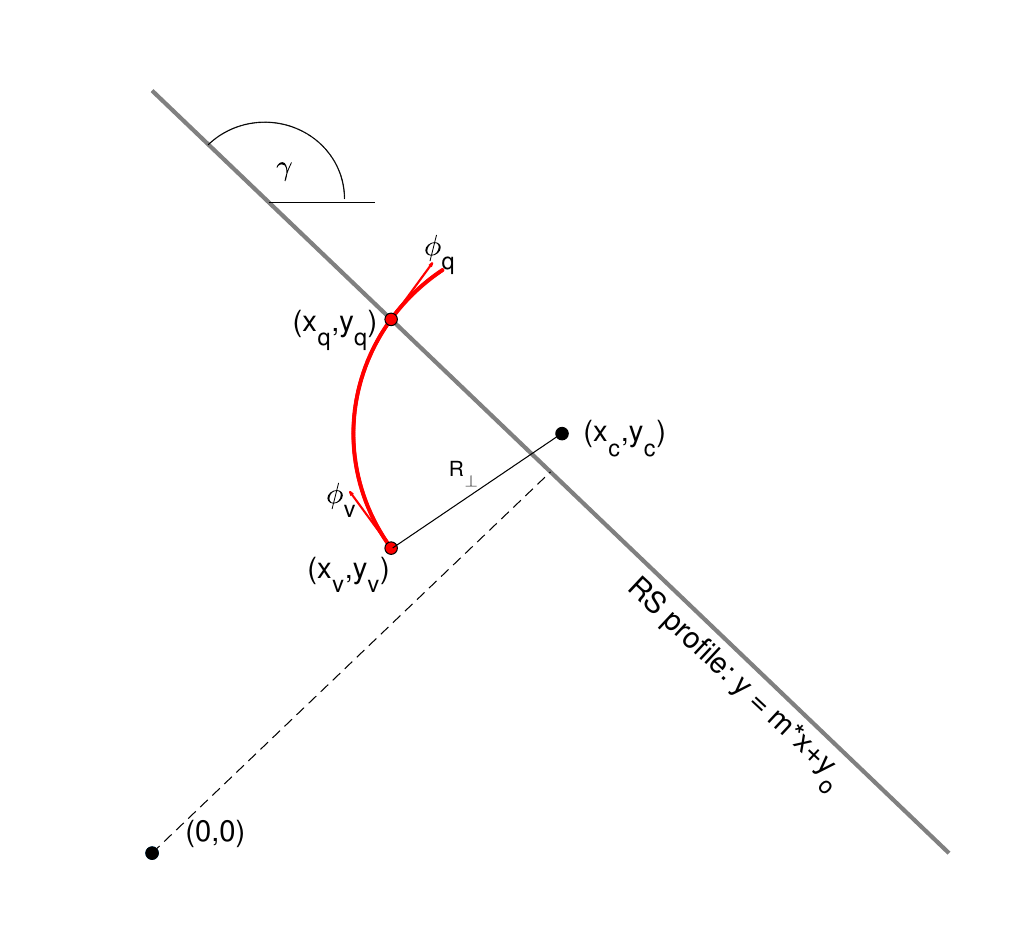}
\caption{A sketch representing the intersection between a trajectory and the detector RS in the x-y plane. }\label{Intersection}
\end{center}
\end{figure}

\begin{equation}\label{EqCenterPosition2}
\left\lbrace
\begin{aligned}
&(y_q-y_c(\textbf{v}, \textbf{t}, k))^2 + (x_q-x_c(\textbf{v}, \textbf{t}, k))^2 = R_{\perp}^2\, ,\\
&y_{q}=m\,x_{q} + y_o\, ,\\
&m=\text{tg}(\gamma)\, .
\end{aligned}
\right.
\end{equation}

\begin{figure}
\begin{center}
\includegraphics[scale=0.35]{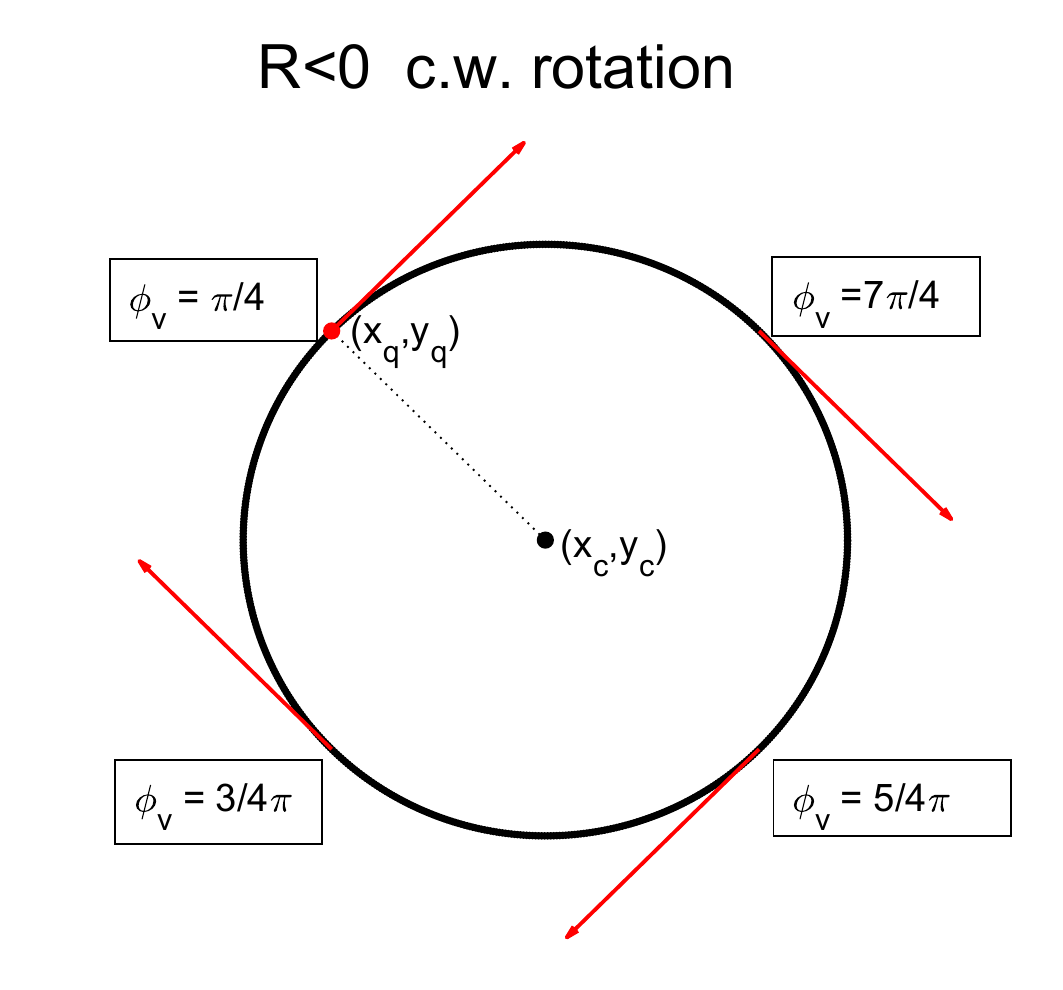}
\caption{Sketch of a track with negative radius, as defined in eq. \ref{Rsign}, projected on the x-y plane. The $\phi_{\text{v}}$ values are obtained by subtracting $\pi/2$ from the phase angle of the radial vector. }\label{phase}
\end{center}
\end{figure}

In the previous equation, $\gamma$ is the angle of the detector ladder with respect to the global $\hat{x}$ axis. The parameter $y_o$ can be calculated by using $y_{\text{meas}} = tg(\gamma)x_{\text{meas}} + y_o$. In conclusion, the solutions of the system of equations \ref{EqCenterPosition2} can be written as:

\begin{equation}\label{solxq}
\begin{array}{l}
x_q=\left(\begin{array}{c}
-\frac{\mathrm{y_o}-\frac{\mathrm{y_o}+m\,\mathrm{x_v}+\sigma_1 +m^2 \,\mathrm{y_v}+\sigma_2 -\sigma_3 }{m^2 +1}}{m}\\
-\frac{\mathrm{y_o}-\frac{\mathrm{y_o}+m\,\mathrm{x_v}-\sigma_1 +m^2 \,\mathrm{y_v}+\sigma_2 -\sigma_3 }{m^2 +1}}{m}
\end{array}\right)\\
\end{array}
\end{equation}
and
\begin{equation}\label{solyq}
y_q=\left(\begin{array}{c}
\frac{\mathrm{y_o}+m\,\mathrm{x_v}+\sigma_1 +m^2 \,\mathrm{y_v}+\sigma_2 -\sigma_3 }{m^2 +1}\\
\frac{\mathrm{y_o}+m\,\mathrm{x_v}-\sigma_1 +m^2 \,\mathrm{y_v}+\sigma_2 -\sigma_3 }{m^2 +1}
\end{array}\right)\, ,\\
\mathrm{}\\
\end{equation}
where
\begin{equation}
\begin{array}{l}
\mathrm{}\\
\;\;\sigma_1 =m\,\left[ -R^2 \,m^2 \,{\mathrm{cos}\left(\mathrm{\lambda_v}\right)}^2 \,{\mathrm{sin}\left(\mathrm{\phi_v}\right)}^2 +R^2 \,m^2 \,{\mathrm{cos}\left(\mathrm{\lambda_v}\right)}^2 \right. \\
\left.-2\,R^2 \,m\,{\mathrm{cos}\left(\mathrm{\lambda_v}\right)}^2 \,\mathrm{cos}\left(\mathrm{\phi_v}\right)\,\mathrm{sin}\left(\mathrm{\phi_v}\right)-R^2 \,{\mathrm{cos}\left(\mathrm{\lambda_v}\right)}^2 \,{\mathrm{cos}\left(\mathrm{\phi_v}\right)}^2 \right. \\ 
\left.+R^2 \,{\mathrm{cos}\left(\mathrm{\lambda_v}\right)}^2 +2\,R\,m^2 \,\mathrm{x_v}\,\mathrm{cos}\left(\mathrm{\lambda_v}\right)\,\mathrm{sin}\left(\mathrm{\phi_v}\right)\right.\\
\left. +2\,R\,m\,\mathrm{x_v}\,\mathrm{cos}\left(\mathrm{\lambda_v}\right)\,\mathrm{cos}\left(\mathrm{\phi_v}\right) +2\,R\,m\,\mathrm{y_o}\,\mathrm{cos}\left(\mathrm{\lambda_v}\right)\,\mathrm{sin}\left(\mathrm{\phi_v}\right)\right. \\
\left. -2\,R\,m\,\mathrm{y_v}\,\mathrm{cos}\left(\mathrm{\lambda_v}\right)\,\mathrm{sin}\left(\mathrm{\phi_v}\right)+2\,R\,\mathrm{y_o}\,\mathrm{cos}\left(\mathrm{\lambda_v}\right)\,\mathrm{cos}\left(\mathrm{\phi_v}\right)\right. \\
\left. -2\,R\,\mathrm{y_v}\,\mathrm{cos}\left(\mathrm{\lambda_v}\right)\,\mathrm{cos}\left(\mathrm{\phi_v}\right)-m^2 \,{\mathrm{x_v}}^2 -2\,m\,\mathrm{x_v}\,\mathrm{y_o} \right. \\
\left. + 2\,m\,\mathrm{x_v}\,\mathrm{y_v}-{\mathrm{y_o}}^2 +2\,\mathrm{y_o}\,\mathrm{y_v}-{\mathrm{y_v}}^2 \right]^{1/2} \\
\mathrm{} \\
\;\;\sigma_2 =R\,m^2 \,\mathrm{cos}\left(\mathrm{\lambda_v}\right)\,\mathrm{cos}\left(\mathrm{\phi_v}\right)\\
\mathrm{}\\
\;\;\sigma_3 =R\,m\,\mathrm{cos}\left(\mathrm{\lambda_v}\right)\,\mathrm{sin}\left(\mathrm{\phi_v}\right)\, .
\end{array}
\end{equation}
A choice between the two solutions in equations \ref{solxq} and \ref{solyq} can be made by accounting for the track direction of motion and the vertex position relative to the hit. 

For what is concerned with $\phi_{\text{v}}$, the following expression can be written, see figure \ref{phase}:
\begin{equation}
\phi_q = \text{atan}\left( \frac{y_q-y_c(x_\text{v},y_\text{v},\phi_\text{v},\lambda_\text{v},k_\text{v})}{x_q - x_c(x_\text{v},y_\text{v},\phi_\text{v},\lambda_\text{v},k_\text{v})}\right) + \text{sign}(R)\pi/2 \, . 
\end{equation}

\subsubsection*{Propagation along the beam axis}
The propagation of helical trajectories along the $\hat{z}$ axis is characterized by the following equations \cite{VALENTAN2009728}:
\begin{equation}\label{TransportZ}
\left\lbrace
\begin{aligned}
&\lambda_q = \lambda_\text{v}\, ,\\
&z_q = z_\text{v} + R_{\perp}\,\text{tan}(\lambda_\text{v})\left( \phi_q - \phi_\text{v}\right)\,  .
\end{aligned}
\right.
\end{equation}

\bibliographystyle{elsarticle-num-names} 
\bibliography{Biblio}

\end{document}